%% file: manuscript.tex
\documentclass[preprint]{aastex631}

\usepackage[colorinlistoftodos,textwidth=0cm]{todonotes}
\presetkeys{todonotes}{inline}{} 

\newcommand\ed[1]{#1}  

\usepackage[acronym]{glossaries}
\makeglossaries

\usepackage{import}
\usepackage{multirow}  
\usepackage{amsmath}   
\usepackage{bm}        
\usepackage{ragged2e}  
\let\tablenum\relax
\usepackage[binary-units=true]{siunitx}[=v2]
\DeclareSIUnit[quantity-product = {}]\frames{\text{frames}}

\graphicspath{{./}{images/}{videos/}}

\newcolumntype{Y}[1]{>{\RaggedRight\let\newline\\\arraybackslash\hspace{0pt}}p{#1}}

\begin{document}

\import{sections/}{0_title.tex}

\import{sections/}{0_glossary.tex}

\import{sections/}{0_abstract.tex}

\import{sections/}{1_introduction.tex}

\import{sections/}{2_owls.tex}

\import{sections/}{3_osia.tex}

\import{sections/}{4_data.tex}

\import{sections/}{5_results.tex}


\import{sections/}{6_infusion.tex}

\section{Funding}
The research was carried out at the Jet Propulsion Laboratory, California Institute of Technology, under a contract with the National Aeronautics and Space Administration (80NM0018D0004).

\section{Acknowledgments}

This research occurred at the Jet Propulsion Laboratory, California Institute of Technology, under a contract with the National Aeronautics and Space Administration. We acknowledge the entire \acrshort{owls} team, especially Andrew Berg, Michael Starch, Santos ``Felipe'' Fregoso, Aaron Noell, Gene Serabyn, \ed{Emily Dunkel, Shawn Anderson, Zaki Hasnain, Ravi Kiran, Peyman Tavallali,} and Marc Foote. We would also like to acknowledge Mae Dubay, Nikki Johnston, and Max Riekeles \ed{as well as the two anonymous reviewers for their insights}. Data annotation for this project was provided by contractors Aman Kumar, Sonali Jain, Aman Sajwan, Bittu Kumar, and Waseem Khan through Labelbox, Inc.

© 2023. All rights reserved.

\section{Disclosures}
The authors declare no conflicts of interest.

\bibliography{bibliography}
\bibliographystyle{aasjournal}

\import{sections/}{7_appendix.tex}

\end{document}

%% file: sections/0_title.tex
\shorttitle{Onboard Science Autonomy for Microscopic Life Detection}

\shortauthors{Wronkiewicz et al.}

\title{Onboard Science Instrument Autonomy for the Detection of Microscopy Biosignatures on the Ocean Worlds Life Surveyor}

\author[0000-0002-6521-3256]{Mark Wronkiewicz}
\affiliation{Jet Propulsion Laboratory, California Institute of Technology, Pasadena, CA, USA}

\author[0000-0002-2838-2878]{Jake Lee}
\affiliation{Jet Propulsion Laboratory, California Institute of Technology, Pasadena, CA, USA}

\author[0000-0002-5233-2029]{Lukas Mandrake}
\affiliation{Jet Propulsion Laboratory, California Institute of Technology, Pasadena, CA, USA}

\author[0000-0001-9811-2227]{Jack Lightholder}
\affiliation{Jet Propulsion Laboratory, California Institute of Technology, Pasadena, CA, USA}

\author[0000-0003-1233-2224]{Gary Doran}
\affiliation{Jet Propulsion Laboratory, California Institute of Technology, Pasadena, CA, USA}

\author[0000-0003-2371-8843]{Steffen Mauceri}
\affiliation{Jet Propulsion Laboratory, California Institute of Technology, Pasadena, CA, USA}

\author[0000-0002-1531-3873]{Taewoo Kim}
\affiliation{Jet Propulsion Laboratory, California Institute of Technology, Pasadena, CA, USA}

\author[0000-0002-7615-3885]{Nathan Oborny}
\affiliation{Jet Propulsion Laboratory, California Institute of Technology, Pasadena, CA, USA}

\author{Thomas Schibler}
\affiliation{Jet Propulsion Laboratory, California Institute of Technology, Pasadena, CA, USA}

\author[0000-0001-5258-0076]{Jay Nadeau}
\affiliation{Portland State University, Portland, OR, USA}

\author[0000-0001-5299-6899]{James K. Wallace}
\affiliation{Jet Propulsion Laboratory, California Institute of Technology, Pasadena, CA, USA}

\author[0000-0003-0928-0531]{Eshaan Moorjani}
\affiliation{Jet Propulsion Laboratory, California Institute of Technology, Pasadena, CA, USA}
\affiliation{University of California, Berkeley, Berkeley, CA, USA}

\author[0000-0002-0146-8372]{Chris Lindensmith}
\affiliation{Jet Propulsion Laboratory, California Institute of Technology, Pasadena, CA, USA}

\correspondingauthor{Mark Wronkiewicz}
\email{wronk@jpl.nasa.gov}

\keywords{Biosignatures(2018) -- Astrobiology(74) -- Saturnian satellites(1427) -- Galilean satellites(627) -- Ocean planets(1151)}

%% file: sections/0_glossary.tex
\glsdisablehyper
\newacronym{owls}{OWLS}{Ocean World Life Surveyor}
\newacronym{elvis}{ELVIS}{Extant Life Volumetric Imaging System}
\newacronym{dhm}{DHM}{Digital Holographic Microscope}
\newacronym{flfm}{FLFM}{Fluorescence Light-Field Microscope}
\newacronym{hrfi}{HRFI}{High Resolution Fluorescent Imager}
\newacronym{osia}{OSIA}{Onboard Science Instrument Autonomy}
\newacronym{acme}{ACME}{Autonomous CE-ESI Mass-spectra Examination}
\newacronym{aegis}{AEGIS}{Autonomous Exploration for Gathering Increased Science}
\newacronym{icer}{ICER}{ICER}
\newacronym{asdp}{ASDP}{Autonomous Science Data Product}
\newacronym{sue}{SUE}{Science Utility Estimate}
\newacronym{dd}{DD}{Diversity Descriptor}
\newacronym{det}{DET}{Decision Error Trade-off}
\newacronym{aucroc}{AUC ROC}{Area Under the Receiver Operating Characteristic Curve}
\newacronym{roc}{ROC}{Receiver Operating Characteristic}
\newacronym{dqe}{DQE}{Data Quality Estimate}
\newacronym{conops}{ConOps}{Concept of Operations}
\newacronym{dsn}{DSN}{Deep Space Network}
\newacronym{modis}{MODIS}{Moderate Resolution Imaging Spectroradiometer}
\newacronym{ml}{ML}{Machine Learning}
\newacronym{ms}{MS}{Mass Spectra}
\newacronym{nisar}{NISAR}{NASA-ISRO Synthetic Aperture Radar}
\newacronym{e2e}{E2E}{End to End}
\newacronym{helm}{HELM}{Holographic Examination for Lifelike Motility}
\newacronym{fame}{FAME}{FLFM Autonomous Motility Evaluation}
\newacronym{jewel}{JEWEL}{Joint Examination for Water-based Extant Life}
\newacronym{mmr}{MMR}{Maximum Marginal Relevance}
\newacronym{snr}{SNR}{Signal to Noise Ratio}
\newacronym{lap}{LAP}{Linear Assignment Problem}
\newacronym{bsub}{B.Sub.}{Bacillus subtilis}
\newacronym{toga}{TOGA}{Tuning Optimizing Genetic Algorithm}
\newacronym{cnn}{CNN}{Convolutional Neural Network}
\newacronym{rf}{RF}{Random Forest}
\newacronym{gbt}{GBT}{Gradient Boosted Tree}
\newacronym{svc}{SVC}{Support Vector Classifier}
\newacronym{mhi}{MHI}{Motion History Image}
\newacronym{shap}{SHAP}{SHapley Additive exPlanations}
\newacronym{msd}{MSD}{Mean-Squared Displacement}
\newacronym{var}{VAR}{Vector AutoRegression}
\newacronym{trl}{TRL}{Technology Readiness Level}
\newacronym{hpsc}{HPSC}{High Performance Space Computer}
\newacronym{oom}{OOM}{Order-of-Magnitude}
\newacronym{msl}{MSL}{Mars Science Laboratory}
\newacronym{leo}{LEO}{Low Earth Orbit}
\newacronym{mro}{MRO}{Mars Reconnaissance Orbiter}
\newacronym{gpu}{GPU}{Graphics Processing Unit}
\newacronym{lds}{LDS}{Life Detection Suite}
\newacronym{maven}{MAVEN}{Mars Atmosphere and Volatile Evolution}
\newacronym{tgo}{TGO}{ExoMars Trace Gas Orbiter}
\newacronym{iadt}{IADT}{Inspection, Analysis, Demonstration, and Test}

%% file: sections/0_abstract.tex
\begin{abstract}


The quest to find extraterrestrial life is a critical scientific endeavor with civilization-level implications. Icy moons in our solar system are promising targets for exploration because their liquid oceans make them potential habitats for microscopic life. However, the lack of a precise definition of life poses a fundamental challenge to formulating detection strategies. To increase the chances of unambiguous detection, a suite of complementary instruments must sample multiple independent biosignatures (e.g., composition, motility/behavior, and visible structure). Such an instrument suite could generate 10,000$\times$ more raw data than is possible to transmit from distant ocean worlds like Enceladus or Europa. To address this bandwidth limitation, \acrfull{osia} is an emerging discipline of flight systems capable of evaluating, summarizing, and prioritizing observational instrument data to maximize science return. We describe two \acrshort{osia} implementations developed as part of the \acrfull{owls} prototype instrument suite at the Jet Propulsion Laboratory. The first identifies life-like motion in digital holographic microscopy videos, and the second identifies cellular structure and composition via innate and dye-induced fluorescence. Flight-like requirements and computational constraints were used to lower barriers to infusion, similar to those available on the Mars helicopter, ``Ingenuity.'' We evaluated the \acrshort{osia}'s performance using simulated and laboratory data and conducted a live field test at the hypersaline Mono Lake planetary analog site. Our study demonstrates the potential of \acrshort{osia} for enabling biosignature detection and provides insights and lessons learned for future mission concepts aimed at exploring the outer solar system.

\end{abstract}

%% file: sections/1_introduction.tex




\section{Introduction}

\subsection{The Search for Life}

The search for life beyond Earth is a driving theme in the 2022 Planetary Science Decadal Survey \citep{Decadal_Survey} to address the civilization-level question, ``Are we alone?'' To translate this search into well-defined planetary mission concepts, we draw insight from our current understanding of terrestrial life while accepting the fewest assumptions possible to preserve sensitivity to exotic forms. On Earth, all life requires access to water. Our solar system includes several bodies known or suspected to contain liquid water, including the icy moons of Jupiter and Saturn with their ancient, deep internal oceans \citep{Hendrix_2018}. Some of these ``ocean worlds'' feature plumes of liquid water streaming into space \citep{hansen_2011}, forming natural sampling opportunities for life detection missions. Even Mars, with its comparatively dry environment, may contain lava tubes with subsurface aquifers sufficient to support past or current microbial communities \citep{LEVEILLE_2010}. While the complexity, size scale, and biochemistry of potential extant life are exceedingly difficult to predict, terrestrial environments suggest that our search should begin with simple, microscopic forms. Bacteria and archaea are the most ubiquitous and numerous forms of life on Earth, significantly predating all multi-cellular life and surviving in the widest variety of habitats \citep{Donoghue_2010} including analog environments similar to what may lie within Enceladus and Europa \citep{Marion_2003, europa_mission_concept_2017}. Therefore, recent mission concepts have focused on instruments to detect microscopic life in aqueous environments.

\subsection{Instruments for the Search}

Life detection is a uniquely challenging scientific objective for two reasons. First, there remains considerable disagreement on the fundamental definition of life on Earth \citep{Lovelock_1965}. Second, for any single proposed biosignature, there are abiotic processes that can generate similar, misleading signals. This is exacerbated on other planetary bodies where dominant physical processes may substantially differ from those studied on Earth \citep{Steele_2022}. To address this, life detection missions should include the capability to detect conceptually orthogonal biosignatures that together reduce the likelihood of misinterpretation of biotic and abiotic phenomena.

The Europa lander mission concept introduces a ``biosignature bingo'' template, where multiple biosignature results can be combined to aid in the joint assessment of a given site \citep{europa_mission_concept_2017, europa_mission_paper_2022}. Proposed instruments to inform this process include a surface stereo camera, luminescence microscope, Raman spectrometer, and gas chromatograph-mass spectrometer among others. Similarly, the Enceladus Orbilander mission concept by \cite{orbilander_mission_concept} identifies independent science objectives to be satisfied by a high-resolution mass spectrometer, laser-induced fluorescence, a separate separation-capable mass spectrometer, and an optical microscope. Finally, the \acrfull{owls} instrument suite includes six instruments, including three microscopes, a mass spectrometer, an organic molecules detector, and laser-induced fluorescence \citep{abscicon_OWLS}. Many of these proposed instruments are modeled after those found in modern biological laboratories and can generate gigabytes of data per observation. While such data volumes are routinely accommodated in a laboratory setting, the need for space missions to communicate all findings across vast interplanetary distances and through over-subscribed resources like the Deep Space Network makes communication bandwidth a primary bottleneck for planetary exploration. Put simply, the compelling detection of extraterrestrial life may require over 10,000 times more raw data than is transmissible by a space mission.

\subsection{Data Bandwidth Limitations at Interplanetary Distances}

The physical limitations of data transmission for planetary missions --- known as the ``bandwidth barrier'' --- directly hinders the search for extant life \citep{Castano_2007_OASIS, Theiling2022_science_autonomy} and was identified as a major challenge in the recent Planetary Science Decadal Survey \citep{Decadal_Survey}. This limitation results primarily from the inverse square law that governs electromagnetic propagation. The recent mission concepts for the Enceladus Orbilander and Europa Lander estimated downlink transmission rates of \SI{34}{\kilo\bit\per\second} and \SI{48}{\kilo\bit\per\second}, respectively \citep{orbilander_mission_concept, europa_mission_paper_2022}. This is less than the \SI{56}{\kilo\bit\per\second} achievable with dial-up internet and would require roughly an hour to transmit a single, uncompressed 12 megapixel image from a modern phone camera. At the same time, life detection mission concepts are pushing the boundaries on large data volume instruments. For just one of the microscopic imagers in the \acrshort{owls} instrument suite, a one minute observation generates as much data as the entire Enceladus Orbilander's surface mission data budget \citep{orbilander_mission_concept} and approximately 40 times that of Europa Lander \citep{europa_mission_concept_2017}. Beyond \ed{the physical data rate limitation}, all deep space missions must also rely on the already over-committed Deep Space Network, further limiting communications opportunities and returned data volume \citep{hackett_2018}. While technology upgrades such as optical communication are underway, these are anticipated to result in at most a 40 times improvement \citep{Deutsch_2020}. Therefore, \ed{novel mission strategies are} required to accommodate high data volume instruments in the planetary context.


Missions currently accommodate the bandwidth barrier by limiting the number of observations, but this approach is poorly suited to the search for life. In this strategy of ``take only what raw data can be returned,'' no observations are requested beyond the available downlink or other limitations (e.g., available power, thermal management, onboard data storage, and competing instrument schedules). For example, \ed{the \ed{Orbilander} concept proposes making up to 7 observations with most life detection instruments during the 176-day primary science phase. This would total $\sim$\SI{1}{\giga\byte} of raw data when including the large data volume nanopore sequencer (or $\sim$\SI{0.29}{\giga\byte} without it; \cite{orbilander_mission_concept})}. This strategy, in the context of traceable science requirements and budget minimization, often drives mission design: why engineer a spacecraft capable of capturing more data than can be returned? While this prioritizes efficient usage of scarce planetary exploration funding, it can also hinder statistically robust sampling of a remote environment and any associated science conclusions. In the MRO/HiRISE example, observations covering only 3-4\% of the surface of Mars were returned between 2006 and 2022, leaving most of the planet unexplored at higher resolution \citep{McEwen_2010, McEwen_2022}. In the case of life detection, this strategy is distinctly problematic for two reasons. First, compelling biosignatures may be relatively rare in collected observations. Second, for every observation containing a strong biosignature, many more will be required to statistically characterize that signal, contextualize it against a heterogeneous background, explore and falsify abiotic interpretations, and inform a process-level understanding of its origin. Low sample numbers incur the catastrophic risk that exotic life is encountered but not captured in instrument observations or that a lone detection cannot be defensibly substantiated \citep{Levin2016_Viking}.

The second method currently used by missions to address limited downlink bandwidth is through data compression techniques, but these cannot produce the compression factors needed without heavy scientific cost. General purpose data compression methods (such as JPEG, MP3, and MP4) and specialized methods for mission operations such as ICER \citep{kiely_2003} offer tunable, lossy compression of at most $10-50\times$ with minimal apparent degradation. For some space applications such as hazard cameras, landing footage, and contextual panoramas, these techniques are useful means to reduce downlink strain. However, while all of these methods were designed to minimize the perception of degradation, they still introduce compression artifacts that can distort and obscure scientific conclusions, limiting their safe application to very low compression ratios \citep{KERNER_2018109}. Likewise, while generic, lossless compression algorithms also exist that remove redundancy, such as PKZIP, they do not result in significant gains when applied to raw science observations. To achieve the four orders of magnitude reduction needed to enable high-volume instruments while preserving valid scientific conclusions, generic compression algorithms will play a limited role. Instead, we require a solution that is driven by a mission's specific science goals and ensures that the most scientifically informative evidence is returned to Earth. 



\subsection{Managing the Bandwidth Barrier with OSIA} \label{sec:managing_bandwidth}

\acrfull{osia} is a unique sub-field of the broader autonomy pursuit focused on science observation content analysis and empowering mission science teams. \acrshort{osia} seeks to maximize a mission's scientific return in the presence of harsh bandwidth constraints relative to instrument data volumes, limited communications opportunities, rare or transient observations of interest, or unanticipated environments. It is comprised of two broad onboard capabilities: observation summarization and data prioritization. Summarization encompasses the capabilities to recognize, characterize, and extract meaningful information from raw observational data. Summarization algorithms must support as similar scientific conclusions as possible while also reducing the data volume of raw observations by several orders of magnitude. Prioritization provides an ordering of summarized or raw observations based on their contents' scientific relevance and contextual importance to the current mission. It can function both within a single instrument's observational record, as well as in the larger context between multiple instruments. A given mission concept may benefit purely from summarization when raw observation contents can be well modeled and efficiently extracted, purely from prioritization when rare but recognizable signals are of primary concern, or from their joint application in more complex \ed{situations}. In all cases, \acrshort{osia} cannot and should not claim to reach scientific conclusions or advance science itself, but rather remain focused on accelerating and extending the mission science team's understanding and control over observation acquisition and downlink. 

\subsection{OSIA in the Broader Autonomy Context}\label{sec:past_osia}

\acrshort{osia} is complementary to, but distinct from, several system-level autonomy applications such as proximity operations \citep{Frontiers_Nesnas}, onboard cruise navigation \citep{bhaskaran2012autonomous}, onboard planning and scheduling \citep{m2020-planner-astra2022}, and automated surface mobility \citep{MSL_auto}. \ed{While} \acrshort{osia} systems are innately focused on recognizing content within science observations, they can provide alerts and guidance to onboard planning and scheduling systems to \ed{trigger} follow-on observations or inform observation site selection. Similarly, \acrshort{osia} can be used to monitor traditionally engineering-focused inputs like hazard avoidance cameras during auto-drive sequences to assess sites of potential scientific interest, to build a record of imagery content, as well as justify decisions to halt for potential hazards or priority science targets. \ed{To} minimize impact on other flight systems, \acrshort{osia} may be bundled with a dedicated computer and storage within a ``smart instrument'' package. 

Previous to the \acrshort{owls} project, several \acrshort{osia} implementations have been demonstrated on late-phase missions with the purpose of building traceable heritage. These early systems tended to be single-purpose and intentionally simple in construction to help overcome perceived risk and present clear value propositions such as biosignature-related mineralogy detection \citep{Mandrake_2012} and opportunistic, transient phenomenon capture \citep{Castano_2008_MER}. The most mature \acrshort{osia} example yet produced is the \acrfull{aegis} system for Mars rovers, operationally deployed on both Curiosity and Perseverance \citep{francis_2015, francis17:aegis}. By enabling autonomous target acquisition for the ChemCam and SuperCam instruments, \acrshort{aegis} provides a systematic baseline site characterization during periods when the rover would otherwise be idle and has increased mission science yield for ChemCam from 256 to 327 collections per sol \citep{francis17:aegis}. In each case, these examples were opportunistically deployed within architectures not originally intended for onboard analysis. Earlier inclusion of \acrshort{osia} during the mission formulation process (\ed{as in} \acrshort{owls}) would enable new missions and scientific objectives, rather than \ed{acting as} \textit{ad hoc} enhancements \ed{to} existing systems.

\subsection{OSIA Driving Requirements}
\label{sec:intro_driving_reqs}

\begin{figure*}[ht]
    \centering
    \includegraphics[width=1.0\textwidth]{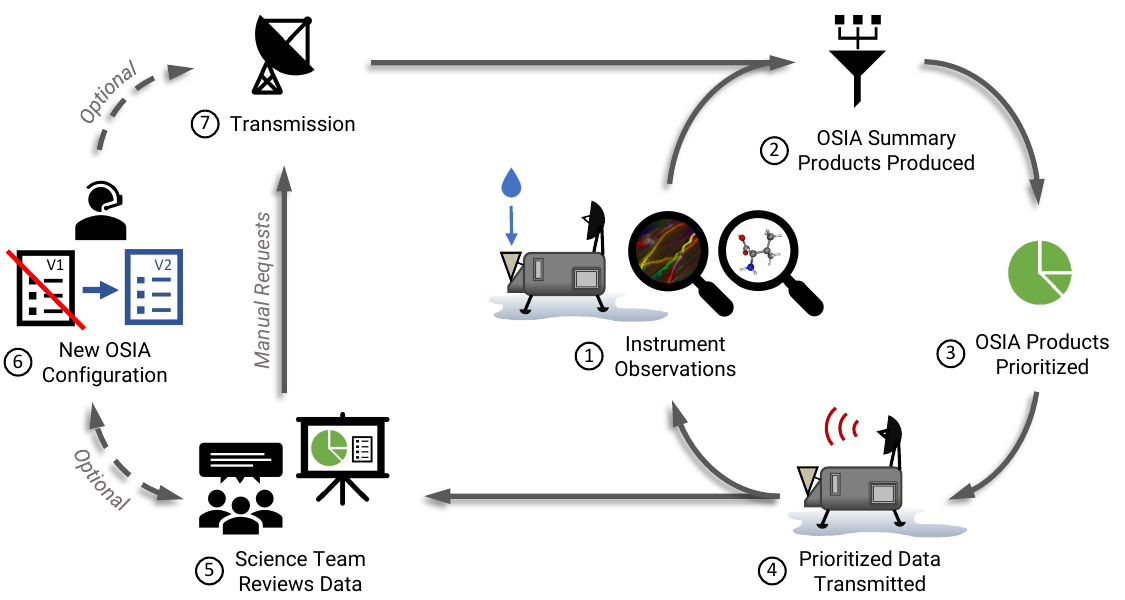}
    \caption{\acrshort{osia} collects, analyzes, summarizes, and prioritizes scientific observations collaboratively with science team guidance and reconfiguration. This simplified illustration depicts a general \acrshort{osia} \acrfull{conops} strategy. Step 1: Ocean world lander collects surface samples and acquires instrument observations. Step 2: \acrshort{osia} assesses and summarizes instrument data according to its current configuration and manual ground requests. Step 3: \acrshort{osia} prioritizes available data products by estimated science value and, optionally, with respect to what has already been returned. Step 4: The highest priority science data products are transmitted to Earth, maximizing downlink capacity. \ed{Lower priority data is retained onboard for potential downlink at a later point in time.} Step 5: Ground science teams review new observations vs. outstanding science questions and determine whether the \acrshort{osia}'s behavior requires redirection. Step 6: If so, scientists and operators generate an updated configuration that is most responsive to the current science intent. Step 7: Any manual data requests and new configurations are transmitted.}
    \label{fig:gotl_diagram}
\end{figure*}

\acrshort{osia} adoption faces challenges due to its operational position between scientists and their raw observations \citep{McGovern_2011}. Figure \ref{fig:gotl_diagram} visualizes this unique \acrfull{conops} strategy. To engender trust with mission stakeholders, we propose that \acrshort{osia} be subject to several novel requirements \citep{Slingerland2022}. 1) To increase transparency and verification, summary data products should include multiple, overlapping extractions from each observation that make different assumptions and use unrelated algorithmic approaches. 2) To support future advances in modeling and analysis, summary products should be accompanied by select, supporting raw data. 3) Observation prioritization should consider explicit science targets of interest, the relative diversity between observations, and the data that has already been returned to the ground. 4) Whether adapting to an evolving science mission focus or changing instrument characteristics, \acrshort{osia} must provide sufficient information about its treatment of observation contents to allow mission operations to recognize the need for and enable \acrshort{osia} reconfiguration. 5) \acrshort{osia} must return sufficient insight to inform manual observation downlink requests, and honor those requests at the highest priority when they occur. 6) Extracted parameters of interest from observations should be accompanied by an estimate of uncertainty to both inform interpretation and capture potential mismatch between the onboard models and observation contents. 

In the following sections, we will specifically consider the \acrshort{owls} instrument suite, evaluate its need for \acrshort{osia} treatment, formulate requirements to drive system design, articulate the chosen implementation, and validate its performance on simulated, laboratory, and field observations.

%% file: sections/2_owls.tex



\section{The Ocean Worlds Life Surveyor}

To empower the search for microbial life on distant planetary bodies, the Jet Propulsion Laboratory has developed the \acrfull{owls}. \acrshort{owls} is a mature, \acrfull{trl} 5, suite of six integrated instruments and onboard software designed to capture and recognize a range of chemical and biological biosignatures in liquid water samples \citep{abscicon_OWLS}. Three microscopes focus on cellular-scale, biological biosignatures within the same sample volume of water: a \acrfull{dhm} to capture video of self-propelled motility at \SI{185}{\nano\meter} scales, a co-observing \acrfull{flfm} to capture video of compounds related to cellular walls, proteins, and nucleic acids at \SI{230}{\nano\meter} scales, and a separate \acrfull{hrfi} that captures complimentary still imagery similar to the \acrshort{flfm} at \SI{60}{\nano\meter} scales \citep{Kim_ELVIS_2020, bedrossian2017digital, serabyn2019development}. \acrshort{owls} also includes three capillary electrophoresis instruments focused on molecular-scale, chemical biosignatures: an ElectroSpray Ionization Mass Spectrometer (ESI-MS) detects biomolecules such as amino acids, the Laser Induced Fluorescence (LIF) instrument analyzes the distribution of chirality in amino acids, and the Capacitively Coupled Contactless Conductivity Detection (C4D) detects organic molecules important to terrestrial metabolism such as tricarboxylic acid and phosphate \citep{mora2022detection, oborny2021radiation, jaramillo2021capillary, willis2022organic}.

The current \acrshort{owls} instrument suite is embodied as an integrated, field-tested engineering prototype designed to evaluate current life detection technology and inform upcoming mission opportunities. During operation, an observation begins as an aqueous sample is delivered to a preparation system that filters out constituents larger than \SIrange{40}{50}{\micro\meter} and adjusts the sample's salinity and particle concentration by increasing or decreasing water content. The prepared sample is then divided between two paths: one is passed directly to the microscopes for biological investigation (as described in the following sections) while the second is treated with heat and pressure to rupture any cellular structures and enable chemical investigation. 

\subsection{Instrument Data Volumes}

As shown in Table \ref{tab:high_data_volume_instruments}, three of the \acrshort{owls} instruments produce observations with sufficient data volume as to benefit from \acrshort{osia} treatment for planetary applications. In this work, we describe \acrshort{osia} developed for the \acrshort{dhm} and \acrshort{flfm} with the associated software available online\footnote{\url{https://github.com/jplmlia/owls-autonomy}} \citep{OWLS-Autonomy-GitHub}. The third instrument, the ESI-MS mass spectrometer to detect small molecules indicative of life, has an equivalent \acrshort{osia} software package called \acrfull{acme} previously described in \cite{Mauceri_2022}.  

\begin{deluxetable}{llrr}
\tablecaption{Typical raw \ed{and lossless compressed} data volumes for a single observation from each \acrshort{owls} instrument.\label{tab:high_data_volume_instruments}}
\tablehead{
\colhead{Instrument} & \colhead{Observation Type} & \colhead{Typical Observation} & \colhead{Lossless Compressed}\\
 & & \colhead{Size (MB)} & \colhead{(ZIP) Size (MB)}
}
\startdata
FLFM   & Video (3D: 2D + time)        & \ed{1,258} & \ed{523}\\
DHM    & Video (3D: 2D + time)        & \ed{1,258} & \ed{1,095}\\
ESI-MS & Ion Count Image (2D)         & \ed{100}    &  \ed{71}\\
HRFI   & Image (2D)                   & \ed{21}    &  \ed{16}\\
C4D    & Time Series (1D)             & \ed{9}     &  \ed{5}\\
LIF    & Time Series (1D)             & \ed{1}     &  \ed{1}\\
\hline
Total &  & \ed{2,647} & 1,711 \\
\enddata
\end{deluxetable}

\begin{figure*}[ht]
    \centering
    \includegraphics[width=1.0\textwidth]{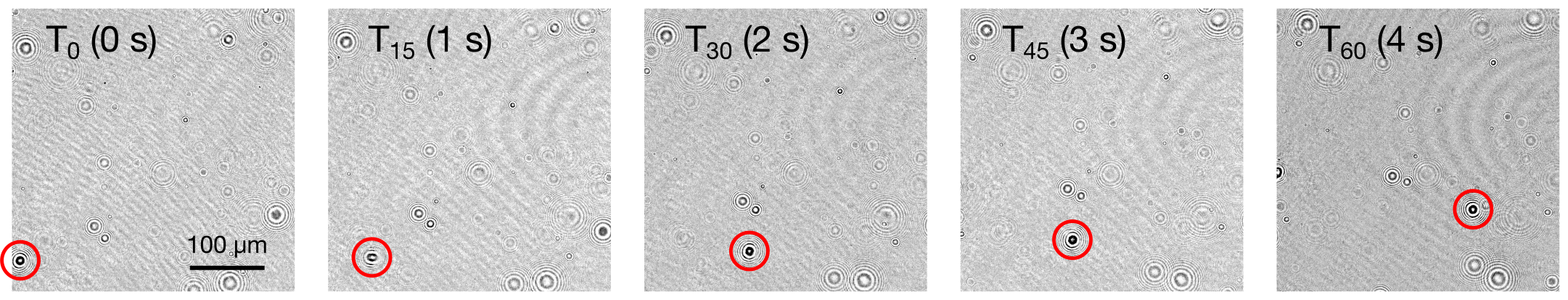}
    \vfill
    \includegraphics[width=1.0\textwidth]{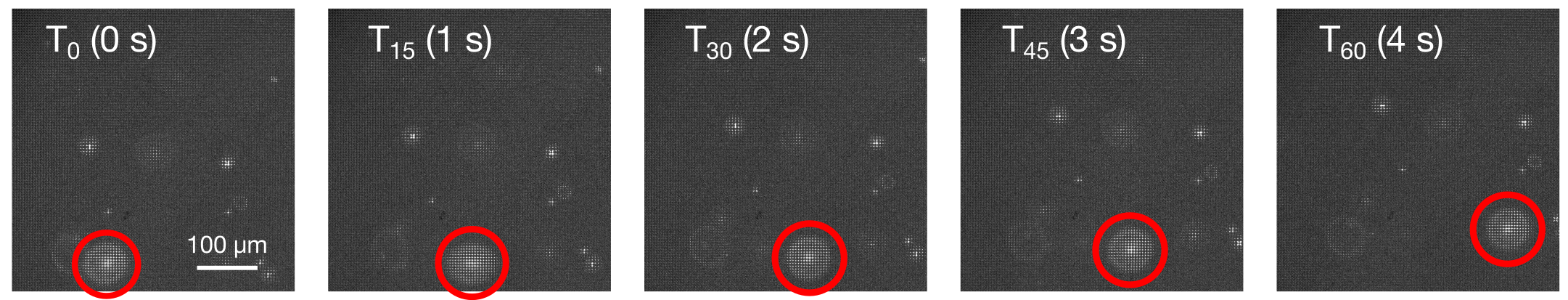}
    \caption{\acrshort{dhm} (top row) and \acrshort{flfm} (bottom row) microscopes simultaneously collect images at 15 frames \si{\per\second} permitting detailed analysis of particle motility- and fluorescence-based biosignatures. These example images show the same motile and fluorescent microorganism in four seconds of raw data for both instruments. The microorganism (circled in red) makes a short rightward movement from T=\SI{1}{\second} to T=\SI{2}{\second} and a much longer burst or movement from T=\SI{3}{\second} to T=\SI{4}{\second}. All other particles slowly drift to the right with the background water flow. The observed Moire pattern in the top row is an artifact resulting from the raw holographic encoding, while the gridded appearance of the organism in the bottom row is an artifact of the light field encoding. Note that both microscopes were still in development when this data was taken, resulting in an alignment offset between the two image series.}
    \label{fig:raw_data_example}
\end{figure*}

\begin{figure}[ht]
    \centering
    \includegraphics[width=0.3\textwidth]{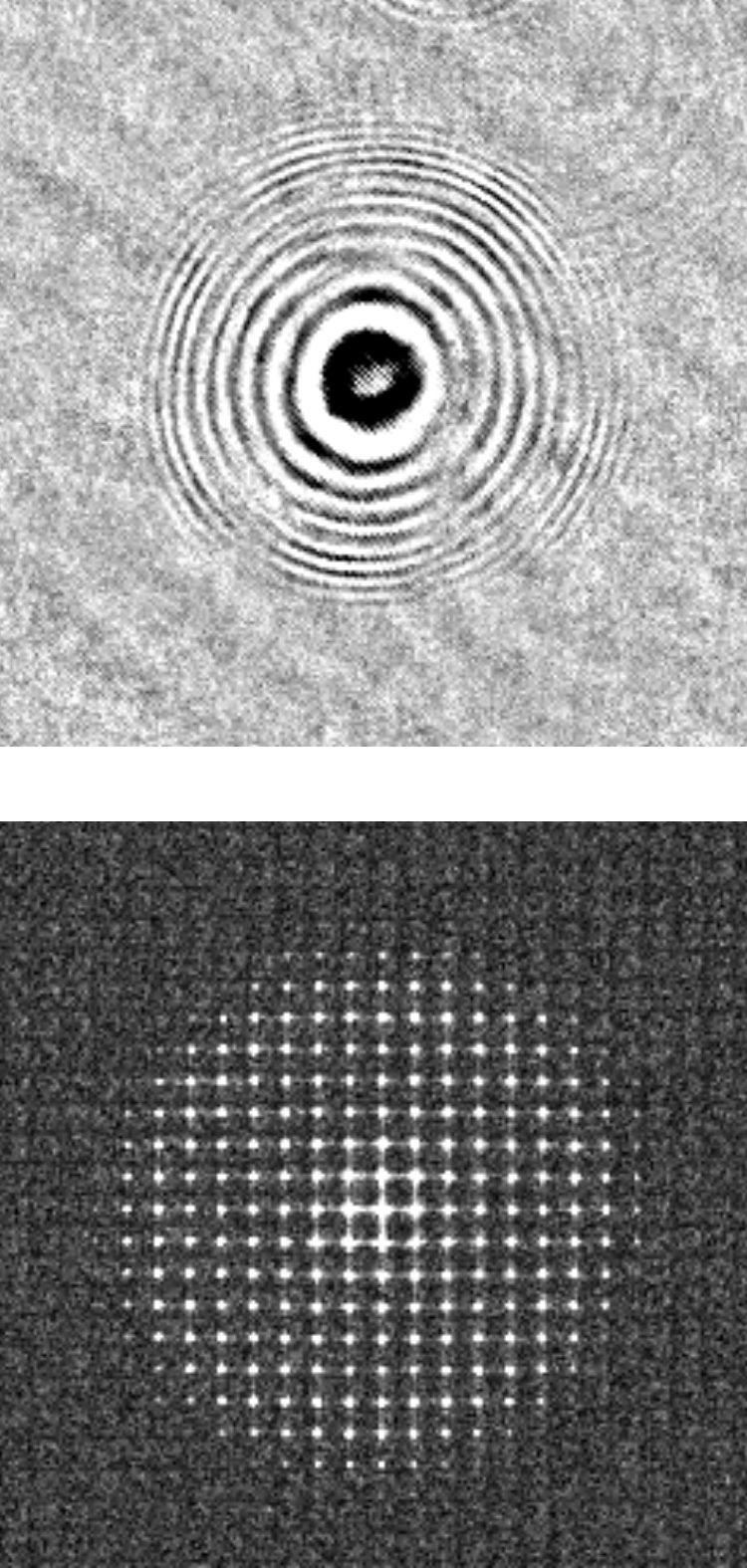}
    \caption{Examples of the same microorganism as it appears in the \acrshort{dhm} (top) and \acrshort{flfm} (bottom) microscopes. The images here are particle crops from T=2s in Figure \ref{fig:raw_data_example}. Particles in the raw \acrshort{dhm} images appear as an Airy pattern that increase in size as the particles become more distant from the central focal plane. Particles in the \acrshort{flfm} appear as a grid of bright spots corresponding to the lenselet array placed in front of the detector. Similarly, the pattern increases in size as the particle is further from the focal plane. While particles may appear distorted to the naked eye, these images contain volumetric information permitting \textit{post hoc} refocusing to any focal plane of interest.}
    \label{fig:raw_portrait_example}
\end{figure}

The \acrshort{dhm} was included on \acrshort{owls} for its ability to image cell-sized particles in a relatively deep sample chamber (using interferometry) \ed{permitting an assessment of} 3D motion. A sample is drawn smoothly through the chamber at \SI{5}{\micro\liter\per\minute} during imaging, continually exposing new particles to inspection. By capturing holographic images at \SI{15}{\frames\per\second}, the \acrshort{dhm} enables the disambiguation of self-propelled particle motility from Brownian motion and fluid dynamics \citep{Wallace_DHM_2015}. Motility is a compelling biosignature that describes the purposeful movement of an organism, such as to search for nutrients, respond to stimuli, or avoid predators \citep{nadeau2016}. Likely because of its impact on the fitness of a species, motility mechanisms have evolved independently in numerous terrestrial organisms \citep{Miyata2020_motility}. In the extraterrestrial context, this biosignature is also fully chemistry- and composition-agnostic. The top row of Figure \ref{fig:raw_data_example} shows example \acrshort{dhm} frames through time containing a motile microorganism, and the top of Figure \ref{fig:raw_portrait_example} shows a close-up of this same microorganism. We implemented an \acrshort{osia} package --- \acrfull{helm} --- to track particle movement within raw \acrshort{dhm} observations and estimate the probability that a given particle \ed{exhibits} motility. \acrshort{helm}'s purpose is to identify and characterize evidence of life-like motion and ensure that data containing these biosignatures is prioritized for transmission to ground teams for further assessment.

The \acrshort{flfm} instrument was included on \acrshort{owls} to capture particle fluorescence induced by an excitation laser \ed{and observes} the same sampling volume as the \acrshort{dhm}. Fluorescence signals of biological interest can originate from innate chemical structures (e.g., chlorophyll) or through fluorescent tags introduced into a liquid sample that selectively bind to specific molecular structures (e.g., lipids in cell-membranes and nucleotides; \cite{Nadeau2008_fluorescence, Serabyn_FLFM_2019}). Both forms of fluorescence permit the measurement of chemical structures known to be important to cellular life on Earth. The bottom row of Figure \ref{fig:raw_data_example} shows example \acrshort{flfm} frames through time containing fluorescent microorganisms, and the bottom of Figure \ref{fig:raw_portrait_example} shows a close-up of a single autofluorescent microorganism. We implemented a second \acrshort{osia} package ---  \acrfull{fame} --- to track and characterize fluorescent particles in raw \acrshort{flfm} observations. Due to the common need for particle tracking, \ed{\acrshort{fame}} shares motility detection algorithms with \acrshort{helm} but includes extensions for fluorescence prioritization. Therefore, \acrshort{fame} has the capability to capture biosignatures related to both cell-like structures as well as motility.

\subsection{Reconstruction vs. Raw Imagery}
\label{sec:owls_volumetric_considerations}

The \acrshort{dhm} and \acrshort{flfm} both produce raw, 2D images that encode the entire 3D sampling volume through time. This encoding produces considerable visual artifacts and distortion when the raw images are directly examined as shown in Figure \ref{fig:raw_data_example} and \ref{fig:raw_portrait_example}. In terrestrial labs with abundant \ed{compute} resources, mathematical reconstruction methods are repeatedly applied to these raw images, extracting focused images at any specified z-plane above/below the central image plane. These reconstructed images are free of encoding artifacts and form a fully volumetric dataset as a function of time. However, image reconstruction is computationally demanding, with full volume reproduction for a single \acrshort{dhm} or \acrshort{flfm} observation requiring hours on a modern computer. This is orders of magnitude beyond what is available for flight hardware in the space exploration context. Thus, both \acrshort{helm} and \acrshort{fame} must operate on the raw \acrshort{dhm} and \acrshort{flfm} imagery directly, treating volumetric encoding artifacts as a noise source. A natural consequence of this decision is that both systems will be highly sensitive to particles near the central plane of the sample chamber where the raw \ed{images} are in focus, with steadily reducing sensitivity as particles move away in depth. In practice, this has little direct negative impact on overall sensitivity. If motile or fluorescent particles are present within the chamber, some of them will be near the central image plane long enough to be detected and characterized. Future missions incorporating \acrshort{helm} or \acrshort{fame} could also opt to include a specialized FPGA-based preprocessor to provide reconstructed imagery \citep{Chen2016LowCF} if additional sensitivity to significantly out-of-focus planes is required.

%% file: sections/3_osia.tex



\section{Onboard Science Instrument Autonomy for OWLS}\label{sec:osia_owls}

\subsection{Defining Mission Success: High-Level Requirements}\label{sec:osia_requirements}

Space missions are developed according to hierarchical requirements that extend from high-level, primary science objectives (level 1) to specific performance requirements for each sub-component (level 4). We defined \ed{field test} requirements early in the formulation of \acrshort{owls} to \ed{focus on infusion into ocean world missions} and lower the barrier for stakeholder trust \citep{Mandrake2022,Slingerland2022}. Given the novelty of these systems, these requirements may also guide future \acrshort{osia}-enabled missions on how to articulate and quantify their own needs, following the guidance of Section \ref{sec:intro_driving_reqs}. The \ed{level 1 through 3} \ed{field test} requirements for the \acrshort{owls} project relevant to its \acrshort{osia} implementation are shown in Tables \ref{tab:reqs12} and \ref{tab:reqs3}, and capture the critical behavior of all \acrshort{owls} \acrshort{osia} independent of implementation details. Level 4 \ed{field test} requirements for \acrshort{helm} and \acrshort{fame}, shown in Table \ref{tab:reqs4}, are described later in Section \ref{sec:requirements} with respect to their detailed implementations.

Our requirements and system architecture are driven first by key assumptions on anticipated image contents. 1) \underline{Non-solitarity} states that if life is present in a properly prepared, concentrated sample, there should be more than one lone organism to recognize. This is supported both by terrestrial analog sites in Antarctica that contain cells at average concentrations of approximately $7.4\times 10^4$ cells/mL, as well as the recommended lower limit of microorganism detection for the Europa Lander study of 100 cells/mL \citep{europa_mission_concept_2017}. This assumption supports the more attainable onboard goal of recognizing and returning some clear, compelling evidence for life, rather than a stringent requirement for the capture of any and all evidence for life across \ed{every observation}. A further assumption of \underline {Ergodicity} states that organisms are expected to freely move through the \ed{3D sample chamber} without preference for any particular location. This allows \acrshort{helm} and \acrshort{fame} to directly analyze the raw, unreconstructed observations despite the loss of sensitivity away from the central z-plane as discussed in Section \ref{sec:owls_volumetric_considerations}. An assumption of \underline{Uncrowded Observations} states that individual organisms will be sufficiently separated within a properly prepared, diluted sample such that their motion track is distinguishable from other particles by about 100 pixels distance in the raw \ed{microscopy observations}. Overly crowded observations containing particles that frequently \ed{cross paths} are difficult to track and assess for biosignatures. \acrshort{helm} and \acrshort{fame} have been designed to detect and provide warnings when observations violate this assumption. Finally, the assumption of being \underline{Well Resolved} states that particles should have sizes between $6\times6$ and $50\times50$ pixels, and their motion should be less than 6 pixels/frame. The initial sample filtration, sample flow rate, \acrshort{dhm} framerate, and microscope focus may be adjusted to ensure these image-based requirements include sensitivity to the intended microorganisms of interest for a given mission use-case. 

\subsection{Autonomous Science Data Products and Prioritization Products} \label{sec:osia_asdp}

A core function of \acrshort{osia} is its ability to summarize raw observations using a standardized set of \acrfullpl{asdp} --- distillations of the raw observation containing only the scientifically relevant information. For \acrshort{helm} and \acrshort{fame}, the \ed{aforementioned assumptions and} requirements (in Tables \ref{tab:reqs3} and \ref{tab:reqs4}) drive their design. Specifically, they: 1) must capture scientifically-relevant \ed{content} using orders of magnitude less data volume, 2) should each assess the observations from a different point of view and goal, using different algorithms and assumptions, 3) should overlap and be mutually reinforcing, such that one \acrshort{asdp} may be used to verify others and falsify alternative conclusions, 4) must enable science team analyses and conclusions similar to raw data return, 5) should include strategically selected raw data that substantiate findings and preserve the potential for future data analyses and modeling, 6) must capture sufficient information to detect and respond to the need for \acrshort{osia} reconfiguration, and finally 7) assess instrument data quality and inform operational monitoring.

Prioritization is realized through the creation of three complementary \acrshortpl{asdp}. The first is the \acrfull{sue}, a positive real number that indicates how similar an observation's contents are to a mission's specified science targets of interest --- here, biosignatures indicating life. The \acrshort{sue} produced by \acrshort{helm} corresponds to the evidence of life-like motility within an observation, while for \acrshort{fame} it must capture both motility and the presence of fluorescence. The second prioritization product is the \acrfull{dqe}, a number that indicates the presence or absence of data quality issues that may impact \acrshort{osia} function. It provides a mechanism to deprioritize observations that fail one or more data quality checks. The third product is the \acrfull{dd}, a vector of several scientifically relevant parameters of interest that meaningfully differentiate one observation's contents from another. The \acrshort{dd} does not compute a single estimate of ``interest'' like the \acrshort{sue}; rather, it enables a second mode of prioritization that orders observations by their relative similarity to, or difference from, other observations. The \acrshort{dd} could be used to request observations that are ``maximally different from what has been already downlinked,'' ``similar to a specific subset of observations,'' or ``best represent the diversity of observations currently onboard.'' As described later in Section \ref{sec:method_prio}, these three \acrshortpl{asdp} enable prioritization (ordering) of observations for downlink. Operationally, \acrshort{helm} and \acrshort{fame} support the dynamic capability to upload \ed{new} specifications for the \acrshortpl{sue}, \acrshortpl{dqe}, and \acrshortpl{dd} to \ed{reconfigure prioritization as desired by science teams}. 

\subsection{The Autonomy Pipeline}

\begin{figure*}[ht]
    \centering
    \includegraphics[width=\textwidth]{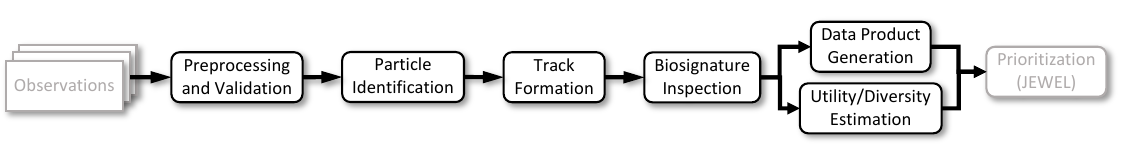}
    \caption{The two \acrshort{osia} algorithms described in this work (\acrshort{helm} and \acrshort{fame}) were designed as a set of modular components to facilitate iterative development. Each module is configurable to support different or changing use cases.}
    \label{fig:system_diagram}
\end{figure*}

The \acrshort{helm} and \acrshort{fame} data processing pipelines search for biosignatures using a multi-step, modular process (Figure \ref{fig:system_diagram}). Both begin by loading microscope observation image frames, optionally preprocessing them to reduce their resolution, and creating background subtracted versions of the frames. Then, the data validation step assesses the observation's data quality, computes the \acrshort{dqe}, and creates some simple contextual \acrshortpl{asdp} (Section \ref{sec:method_preproc_valid}). After that, particles are identified in individual frames and linked across frames (through time) to form particle tracks (Section \ref{sec:method_tracking}). Next, tracks are assessed for biosignatures of interest: \acrshort{helm} identifies signs of motility by extracting track features (Section \ref{sec:method_feature}) and classifying motile movement (Section \ref{sec:method_biosig}), and \acrshort{fame} additionally extracts fluorescence characteristics. Finally, \acrshortpl{asdp} and prioritization products (i.e., the \acrshort{sue} and \acrshort{dd}) are generated for prioritization by \acrfull{jewel} (Section \ref{sec:method_prio}). Each step is modular and configurable for flexibility during the development cycle and mission operations.

\subsubsection{Data Preprocessing and Validation} \label{sec:method_preproc_valid}

The data preprocessing step applies simple parallelized image manipulations to prepare an observation for further analysis. Both \acrshort{dhm} and \acrshort{flfm} data are resized from dimensions of $2048\times2048$ to $1024\times1024$ pixels, which reduces the compute and memory costs of the downstream particle identification and tracking algorithms. The resizing process is parallelizable, and can be configured for even lower resolutions at some cost to particle tracking precision. 

\begin{figure}[ht]
    \centering
    \includegraphics[width=0.4\textwidth]{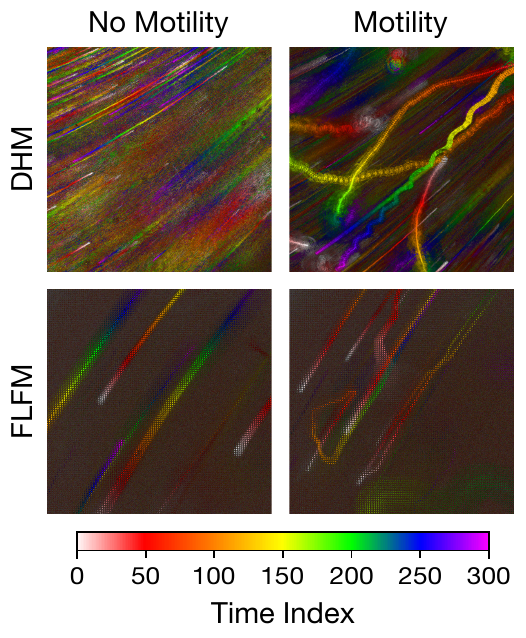}
    \caption{Motion History Images (MHIs) compress microscopy observations into a single, quickly-interpretable image. Here, color indicates the time when each pixel changed maximally. \textbf{Left:} Relatively straight color tracks in DHM (top left) and VFI (bottom left) observations \ed{depict} particles that passively floated through the sample chamber with the background fluid flow. \textbf{Right:} Curving/spiraling tracks in DHM (top right), and reversing tracks in VFI (bottom right) indicate the presence of living, motile cells.}
    \label{fig:mhi_examples}
\end{figure}

The data validation step has two goals: 1) summarize the entire observation to provide contextual information to science teams and 2) execute quality checks to identify potential problems with the instrument or science data. For example, \ed{each \acrfull{mhi} in Figure \ref{fig:mhi_examples}} provides a quickly-interpretable image to understand the quantity and movement characteristics of all particles in an observation. A full list of validation products is described in Table \ref{tab:validation}. Many products were developed in direct response to science needs or data problems encountered during the development of \acrshort{owls}. \ed{For example, the pixel intensity and pixel difference time series products} ensure that observation frames have stable characteristics through time. \ed{Large} pixel differences between frames can indicate instrument vibration, which will negatively affect particle tracking. The previously described \acrshort{dqe} aggregates these pass/fail checks through a weighted summation.

\begin{deluxetable*}{Y{1in} Y{1in} Y{4in}} 
\tablecaption{Data validation products described here are calculated for both instruments unless specified otherwise. The automated checks are used to calculate the \acrshort{dqe}. \label{tab:validation}}
\tablehead{
    \colhead{Name} & \colhead{Type} & \colhead{Justification}
}
\startdata
\acrfull{mhi} & Contextual Product & \ed{Ground teams need a method to quickly understand observations. This single image summarizes} an entire (video) observation \ed{by capturing} the point in time when the maximum change occurred at each pixel.\\ \hline
Median Image & Contextual Product & Single image containing the median value at each pixel location through an entire observation. Used for background subtraction.\\ \hline
Pixel Intensity & Bounded Range Check & The mean pixel intensity of frames should lie within an expected range. Values outside this range could indicate incorrect laser configuration or a blockage in sample flow.\\ \hline
Pixel Difference & Bounded Range Check & The mean frame-to-frame pixel change should lie within an expected range. Excessive pixel differences could indicate instrument vibration or an over-crowded sample.\\ \hline
Estimated Particle Density & Threshold Check & The total number of particles in the field of view should be limited (e.g., through sample dilution) to prevent frequent track crossings (and degradation in tracker performance).\\ \hline
Interframe Interval & Distribution Check & The time interval between consecutive observation frames should be consistent. Unsteady frame rates may indicate the acquisition computer is overloaded and will lead to challenges in assessing particle motion.\\ \hline
2D Image Spectrum (\acrshort{dhm} Only) & Laser Validation Check & The frequency and power of the three \acrshort{dhm} lasers must be properly configured to permit reconstruction. A 2D Fourier spectrum of the observation frames should reflect a known pattern to confirm image reconstruction will be possible by science teams.\\ \hline
\enddata
\end{deluxetable*}

\subsubsection{Particle Identification and Track Formation} \label{sec:method_tracking}

To extract biosignatures, \acrshort{helm} and \acrshort{fame} must first identify and characterize particle motion. Particles are detected in each frame by identifying pixels that are substantially different than the background field (calculated as the median frame across the entire observation). \ed{We} then apply the DBSCAN algorithm \citep{Ester_1996} to find clusters of pixels meeting a specific size threshold, which are deemed particles. \ed{This identification method is agnostic to particle morphology, and tracking directly from the unreconstructed hologram avoids the computationally expensive reconstruction process, which is currently too resource intensive for onboard implementation \citep{marin2018wavelet}.}

The particles are then associated into motion ``tracks'' over time via the \acrfull{lap} tracking algorithm \citep{jaqaman2008robust}. \ed{The LAP tracker first employs frame-to-frame particle linking, assigning nearby particles in consecutive frames into track segments. This step is spatially global but temporally greedy; it makes no assumptions about particle motion characteristics, but it is also brittle to time gaps in particle detection. Therefore, the tracker employs a second gap-closing step, using global optimization to link the starts and ends of track segments into complete tracks. This method is computationally efficient and produces tracks that are robust to short particle occlusions and overlaps. However, extremely crowded samples can overwhelm the tracker by confusing the initial identification or violating the global assumption --- that a particle will be nearest to itself as it moves between frames. If such conditions are expected, increasing the resolution and framerate can improve the efficacy of this tracking method.} 


We formalize the particle tracks produced by this step in preparation for the following sections. An observation of $n$ frames of $p \times p$ pixel resolution results in a set of tracks $\bm{K}$. Each track $k \in \bm{K}$ is a set of particle positions $(x, y)_t$, where $x$ and $y$ are pixel coordinates of the center of the particle, and $t$ is the frame number, representative of time. A track starts and ends at times $t_s$ and $t_e$, and the particle tracking algorithm imposes a minimum on $t_e - t_s$ to filter out spurious false positive identifications. We formally define a track with Equation \ref{eq:trackdef}:

\begin{align}
    k &= \{(x,y)_t \mid 0 \leq (x,y) \leq p, 0 \leq t_s \leq t \leq t_e\leq n \} \label{eq:trackdef}
\end{align}

For all tracks in $\bm{K}$, the spatiotemporal coordinates are stored for further biosignature assessment (i.e., motility and fluorescence) in future processing steps. Track coordinates are also included for transmission as the compact collection of integers \ed{has a relatively small data volume} and permits manual review \ed{of particle motion} by science teams.

\subsubsection{Feature Extraction} \label{sec:method_feature}

Once particles are tracked, \acrshort{helm} and \acrshort{fame} must assess them for signs of motility. To our knowledge, existing systems for motility characterization only evaluate specific species, detect specific patterns (e.g., the run-and-tumble motion of \textit{E. coli}), or are currently too computationally expensive to deploy on spacecraft-ready computers \citep{Son2015, rosser2013novel}. Instead, we developed an onboard system to identify any motion that cannot be explained by Brownian motion and simple fluid dynamics. The \acrshort{osia} achieves this by quantitatively describing each track with a feature vector where every element in the vector is calculated using a different movement metric. This vector is then classified by an \ed{\acrfull{ml}} model to estimate the probability of motility.

Simple motion features, like speed and acceleration (Equations \ref{eq:speed} and \ref{eq:acc} respectively), are calculated to identify particles with changing movement patterns. The mean, standard deviation, and maximum value of these discrete values are included as features.

\begin{align}
    \bm{s} &= \{ \|(x,y)_t - (x,y)_{t-1}\|_2 \mid t_s < t \leq t_e \} \label{eq:speed}\\
    \mu(\bm{s}) &= \text{Mean Speed} \nonumber\\
    \sigma(\bm{s}) &= \text{Standard Deviation of Speed} \nonumber\\
    \max(\bm{s}) &= \text{Maximum Speed} \nonumber\\
    \nonumber\\
    \bm{a} &= \{ s_t - s_{t-1} \mid t_s + 1 < t \leq t_e \} \label{eq:acc}\\
    \mu(\bm{a}) &= \text{Mean Acceleration} \nonumber\\
    \sigma(\bm{a}) &= \text{Standard Deviation of Acceleration} \nonumber\\
    \max(\bm{a}) &= \text{Maximum Acceleration} \nonumber
\end{align}

We also measure a particle's step angle at each frame (the discrete angular velocity; Equation \ref{eq:stepang}), as frequent or large deviations from a straight path suggest motility. Again, the mean, standard deviation, and maximum values are included as features. Here, we omit the conversion of step angle radian values to the range $[-\pi, \pi]$ for simplicity.

\begin{align}
    \bm{\theta} &= \{\arctan\left(\frac{y_t - y_{t-1}}{x_t - x_{t-1}} \right) \mid t_s < t \leq t_e \} \label{eq:stepang}\\
    \Delta \bm{\theta} &= \{ |\theta_t - \theta_{t-1}| \mid t_s + 1 < t \leq t_e \} \nonumber \\
    \mu(\Delta \bm{\theta}) &= \text{Mean Step Angle} \nonumber \\
    \sigma(\Delta \bm{\theta}) &= \text{Standard Deviation of Step Angle} \nonumber \\
    \max(\Delta \bm{\theta}) &= \text{Maximum Step Angle} \nonumber
\end{align}

\ed{We can treat each of the speed, acceleration, and step angle features as a time series} and measure their autocorrelation for any time lag. This quantifies whether a particle exhibits a movement pattern with some periodicity. We generate features with time lags of 15 and 30 frames (corresponding to 1 and 2 seconds, respectively), but note that additional time offsets could be added.

In addition to frame-discrete features, we calculate features describing holistic track movement. The length of the track is measured in pixels (Equation \ref{eq:len}) and the \ed{duration} of the track is measured in frames (Equation \ref{eq:lifetime}). The horizontal, vertical, Euclidean, and angular displacements from the start to end of the track are also measured (Equations \ref{eq:e2ehd}, \ref{eq:e2evd}, \ref{eq:e2enorm}, \ref{eq:e2eang}); since there is often background fluid flow in the sample chamber, the direction of the total displacement could be indicative of a particle moving against this flow. Sinuosity, the ratio between the track length and the total displacement, quantifies movement inefficiency, as motile particles may appear to meander while exploring for nutrients or responding to stimuli (Equation \ref{eq:sin}). Finally, the \acrfull{msd} slope (as implemented in \cite{Manzo_2015}) is used to distinguish Brownian motion from other types of motion.

\begin{align}
    \text{Track Length} &= len = \sum_{t=t_s+1}^{t_e}{s_t}  \label{eq:len}\\
    \text{Track \ed{Duration}} &= t_e - t_s \label{eq:lifetime}\\
    \text{End-to-end Horiz. Disp.} &= x_{t_e} - x_{t_s} \label{eq:e2ehd}\\
    \text{End-to-end Vert. Disp.} &= y_{t_e} - y_{t_s} \label{eq:e2evd}\\
    \text{End-to-end Euclidean Disp.} &= disp = \|(x,y)_{t_e} - (x,y)_{t_s}\|_2 \label{eq:e2enorm}\\
    \text{End-to-end Disp. Angle} &= \arctan\left(\frac{y_{t_e}-y_{t_s}}{x_{t_e}-x_{t_s}}\right) \label{eq:e2eang}\\
    \text{Sinuosity} &= \frac{len}{disp} \label{eq:sin}\\ 
    \text{MSD Slope} &: \ \text{defined by \cite{Manzo_2015}}
\end{align}

While these features describe each track independently, we also wish to identify tracks $k$ that behave differently from the set of all other tracks in an observation, $\{\mathbf{K} \setminus k\}$. We generate the relative speed feature to capture the ratio between a track's mean speed and the mean speed of all other tracks (Equation \ref{eq:relspeed}). Similarly, the relative step angle features are compared between a track's mean step angle and the mean step angle of all other tracks (Equations \ref{eq:relcos}, \ref{eq:relang}). The mean horizontal and vertical displacements of each track (Equations \ref{eq:mhd} and \ref{eq:mvd}) are also provided as features.

\begin{align}
    \text{Rel. Speed} &= \frac{\mu(s)_k}{\mu(\mu(s)_{\{\bm{K}\setminus k\}})} \label{eq:relspeed}\\
    \text{Mean Horiz. Disp.} &= mdx_{k} = \mu( \{ x_t - x_{t-1} \mid t_s < t \leq t_e \}) \label{eq:mhd}\\
    \text{Mean Vert. Disp.} &= mdy_{k} = \mu( \{ y_t - y_{t-1} \mid t_s < t \leq t_e \} ) \label{eq:mvd}\\
    \text{Rel. Step Angle Cos. Sim.} &= \text{cossim}( (x,y)_{k, t_e} - (x,y)_{k, t_s}, (\mu(mdx_{\{\bm{K}\setminus k\}}), \mu(mdy_{\{\bm{K}\setminus k\}}))) \label{eq:relcos}\\
    \text{Rel. Step Angle Diff.} &= \left| \arctan\left( \frac{y_{k, t_e} - y_{k, t_s}}{x_{k, t_e} - x_{k, t_s}} \right) - \arctan\left(\frac{\mu(mdy_{\{\bm{K}\setminus k\}})}{\mu(mdx_{\{\bm{K}\setminus k\}})}\right) \right| \label{eq:relang}
\end{align}

In total, we compute 23 features to quantify the movement characteristics of each track in preparation for biosignature analysis, but additional features can be added if needed. While these features are not downlinked, they can be recalculated on the ground from downlinked track coordinates. With these track features, \acrshort{helm} and \acrshort{fame} are able to assess tracks for evidence of motility.


\subsubsection{Identifying Motility and Fluorescence Biosignatures} \label{sec:method_biosig}

After calculating track features, \acrshort{helm} and \acrshort{fame} investigate tracks for evidence of motility and fluorescence. For both, a \acrshort{ml} classifier estimates the probability that each track exhibits motile behavior. While the system is agnostic to the \acrshort{ml} model used, \ed{the computational and interpretability requirements (Table \ref{tab:reqs4} Req. L4-7,9)} lead us to deploy classical \acrshort{ml} methods, including \acrfullpl{gbt}, \acrfullpl{rf}, and \acrfullpl{svc}. We describe the implementation and evaluation of our model in Section \ref{sec:results_class}. The posterior probabilities \ed{(i.e., confidences)} produced by these classifiers further inform data downlink prioritization \ed{via the \acrshort{sue} (as }discussed in Section \ref{sec:method_prio}). In a flight scenario, the classification model would be trained on the ground and stored onboard for inference with the possibility to retrain and re-transmit new model parameters throughout the mission. \acrshort{fame} also assesses particle fluorescence by directly analyzing the color profile of tracked particles. It calculates maximum particle fluorescence of each color channel and uses this to inform downstream prioritization through \ed{both the \acrshort{sue} and \acrshort{dd}.} 

\subsubsection{Preparing Downlink Products} \label{sec:method_portrait}

Finally, we excise small image crops (or ``portraits'') from the original, full resolution \acrshort{dhm} or \acrshort{flfm} frames for tracks of interest. These portraits can then be reconstructed on the ground to retrieve a volumetric image of each particle \citep{mckeithen2021fast} as demonstrated in Section \ref{sec:results_portraits}. While these represent important \acrshortpl{asdp} as they are portions of the raw data, they also use  considerable data bandwidth. Therefore, users can configure the size of particle portraits, the number of desired portraits per track, and optionally limit transmissions to only include \ed{portraits for} motile tracks in \acrshort{dhm} data.

The final output of the \acrshort{osia} processing pipeline is a list of \acrshort{asdp} bundles from each observation. Each bundle includes: 1) data validation and contextual products, 2) particle tracks and results from their respective biosignature investigations, and 3) original-resolution portraits of scientifically interesting particles. 

\subsubsection{Prioritizing Autonomous Science Data Products for Downlink} \label{sec:method_prio}

As described in Section \ref{sec:osia_asdp}, \acrshort{helm} and \acrshort{fame} compute \ed{a} \acrshort{sue}, \acrshort{dd}, and \acrshort{dqe} for each processed observation. \ed{\acrshort{helm}'s \acrshort{sue} is defined to assign} high science importance \ed{to observations with tracks that are long in duration and have high motility probabilities}. \ed{Motile particles that stay in the field of view for a long duration present the best opportunities for further evaluation.} First, we score each track by multiplying each track's motility probability\ed{, $P(\text{motile} \mid k_i)$, by its \ed{duration}, $\lvert k_i \rvert$}. \ed{Next}, we sum the \ed{top five} scores, then normalize them to $[0,1]$ by dividing by the ideal scenario: five \ed{full-duration} tracks with $1.0$ motility probability. This \acrshort{sue} definition is expressed in Equation \ref{eq:helmsue}, where $k$ is a track vector as previously defined, $n$ is the \ed{duration} of the entire observation, and the summation assumes the tracks are sorted by their motility probability. We discuss the efficacy of this \acrshort{sue} definition in Section \ref{sec:results_summarization}. 

\begin{align} \label{eq:helmsue}
    SUE=\frac{\sum_{i=1}^5 P(\text{motile} \mid k_i) \cdot \lvert k_i \rvert}{5n}
\end{align}

\ed{\acrshort{fame}} computes its \acrshortpl{sue} by calculating the median of tracks' fluorescence intensities. Since a particle must fluoresce to be observed in \acrshort{flfm} data, the median places scientific value on observations with a population of strongly fluorescing particles.

\ed{To quantify diversity,} \acrshort{helm} includes \ed{particle size, speed, and displacement features in its \acrshort{dd} in order to incorporate particle morphology and overall particle movement information into the prioritization}. We discuss the efficacy of this definition in differentiating types of observations in \ed{Section \ref{sec:results_dd}}. \ed{\acrshort{fame}} defines its \acrshort{dd} similarly but also includes the pixel intensities across three image bands to capture the fluorescence profile induced by the binding of different fluorescent tags (or autofluorescence). Refer to Section \ref{sec:method_preproc_valid} for the calculation of the \acrshort{dqe}.

To balance utility, diversity, and data quality, a system called \acrfull{jewel} was developed to prioritize \acrshortpl{asdp} from \ed{a} set of processed observations for downlink \citep{doran2021jewel}. Intuitively, \acrshort{jewel} seeks to select \acrshortpl{asdp} for observations that balance strong evidence of biosignatures and are also diverse compared to previously transmitted observations' \acrshortpl{asdp}. If all instruments calculate the aforementioned prioritization metrics, \acrshort{jewel} can generate a single prioritization queue for a multi-instrument platform like \acrshort{owls}. The prioritization algorithm is based on the \acrlong{mmr} algorithm \citep{carbonell98mmr}, which iteratively selects observations that maximize the additional science utility after applying a ``discount factor'' based on the most similar observation from those already downlinked. \ed{It estimates the similarity of observations by calculating the Gaussian similarity between} pairs of \acrshortpl{dd} \ed{(described more in Section \ref{sec:results_dd})} of each candidate observation and the most similar previously downlinked observation. More formally, from the similarity metric, a \emph{diversity factor} ($\mathrm{df}_{i}$) is computed for the $i^{\mathrm{th}}$ observation as follows:
\begin{equation}
    \label{eq:diversity-factor}
    \mathrm{df}_{i} = (1 - \alpha) + \alpha\left(1 - \max_{j}\mathrm{sim}\left(\mathrm{DD}_{i}, \mathrm{DD}_{j}\right)\right),
\end{equation}
where $\mathrm{DD}_{j}$ are the \acrshortpl{dd} for all previously downlinked data products. The $\alpha\in\left[0,1\right]$ parameter provides a mechanism to control the degree to which diversity-based discounting is applied as opposed to using the initial \acrshort{sue} values. When $\alpha = 0$, the diversity factor is always $1.0$ and no discount is applied (e.g., when observations with the strongest identified biosignatures are desired). On the other hand, when $\alpha = 1$, the similarity-based discount factor is fully applied (e.g., when an equal balance of utility and diversity is desired). After each iteration of selecting a product for downlink, the marginal SUE values are recalculated according to Equation \ref{eq:sue_marginal} to select the next observation for downlink:
\begin{equation}
    \label{eq:sue_marginal}
    \mathrm{SUE}_{i}^{\mathrm{marginal}} =  \mathrm{SUE}_{i} \times \mathrm{df}_{i} \times \mathrm{DQE}_i
\end{equation}
\acrshort{jewel} stores a simple onboard manifest of each observation's prioritization metrics and their downlink status to permit future prioritization for an arbitrary number of downlink events.

To engender trust with scientists and mission operators, \acrshort{jewel} provides two methods of modifying or overriding the \acrshort{osia}'s prioritization decisions. First, \acrshort{jewel} replicates the commonly used concept of ``priority bins'' for downlink prioritization, where high priority \acrshortpl{asdp} are transmitted before moving to the next bin. Operators can use \textit{a priori} knowledge to direct observations \acrshortpl{asdp} to specific bins if their importance is known in advance. Mission operators can also use this mechanism to ensure that some non-zero amount of high priority \acrshortpl{asdp} are transmitted for each instrument regardless of the global (across-instrument) prioritization scheme. Second, operators have the ability to manually override \acrshort{sue} values or move \acrshortpl{asdp} to different bins during ground-in-the-loop commanding opportunities. This may be necessary, for example, if science teams believe an observation's \acrshortpl{asdp} are more, or less, valuable than the \acrshort{osia}'s original estimation. Note that \acrshort{jewel} also permits different per-bin configurations if a sophisticated prioritization scheme is desired. 

While a full operator interface is outside the scope of this work, \acrshort{jewel} generates an interactive ground report to expose \acrshortpl{asdp} and prioritization decisions to ground teams. The visualization displays the cumulative \acrshort{sue} of all downlinked products, a dimensionality-reduced representation of \acrshortpl{dd}, each observation's \acrshortpl{dqe}, and a subset of \acrshortpl{asdp}. The visualization was developed for the Mono Lake field campaign to quickly inform scientists of any identified biosignatures and explain the \acrshort{osia}'s summarization and prioritization decisions.

\subsection{HELM and FAME Field Test Requirements} \label{sec:requirements}

\ed{As discussed in Section \ref{sec:osia_requirements}, we defined a set of field test requirements specific to the \acrshort{osia} described to motivate our work and quantify success at the recently completed \acrshort{owls} field campaign at Mono Lake, CA. (Appendix \ref{apx:reqs}). While formulating a full mission architecture and the associated flight requirements is beyond the scope of this work, these notional autonomy-focused requirements (Table \ref{tab:reqs4}) ensured alignment with the \acrshort{owls} project scientists and instrument developers. Note that these notional requirements are identical for both \acrshort{helm} and \acrshort{fame}} save for extensions related to \acrshort{fame}'s additional capability to evaluate fluorescence biosignatures \ed{(Req. L4-4). In general, the requirements do not motivate} perfect recognition of all biosignatures, but rather ensure that at least one downlink opportunity contains compelling evidence of life. Transmitting false positives has no intrinsic cost, so long as true positives are also returned \ed{(Req. L4-1,3)}. The return of ``near miss'' examples may also be of scientific interest to provide context for strong evidence of life, and even ``clear mistakes'' may inform background population studies and potential \acrshort{osia} improvements or reconfiguration. However, if these false positives crowd out true positives from a downlink opportunity, there is effectively an infinite cost and risk of mission failure \ed{(Req. L4-2)}. Thus, \acrshort{helm} and \acrshort{fame} must quantify the scientific value of any given observation relative to the population of other available observations given the downlink budget \ed{(Req. L4-5,6)}. Beyond prioritizing biosignatures, background and contextual data products are also required to summarize non-prioritized observations in a highly data-efficient manner \ed{(Req. L4-8,9)}. These both inform mission science teams of what might remain onboard and substantiate the high-priority findings. \ed{Finally, a computation requirement ensures the timeliness of \acrshort{osia} in order to support field scientists in real time (Req. L4-7)}. We present these requirements as demonstrations and to seed discussions with future mission opportunities. The specific, quantified values within each requirement will need updating based on a given mission concept's specific communication budget and risk profile.

%% file: sections/4_data.tex



\section{Data} \label{sec:data}

Data for the development and characterization of \acrshort{owls} \acrshort{osia} was assembled from multiple, evolving instrument versions in parallel with the instrument development. Problematic observations were used to develop data validation products as discussed in Section \ref{sec:method_preproc_valid}, while high-quality observations were curated into a quantitative evaluation test set. In addition, we developed a \acrshort{dhm} data simulator to generate synthetic observations with greater control of instrument and sample properties for sensitivity analyses. Finally, we participated in a field campaign to test the \acrshort{osia} outside a laboratory setting. Table \ref{tab:datasets} summarizes these three datasets. For consistency of reporting and evaluation, we standardized all observations to 300 video frames representing 20 seconds of microscopic footage with an average raw data volume of \ed{\SI{1.26}{\giga\byte}}. In practice, a mission would be able to dynamically select an observation's frame rate and duration as driven by the current science focus.

\begin{deluxetable}{lrrrrrr}
\tablecaption{Summary of curated \acrshort{dhm} and \acrshort{flfm} datasets.\label{tab:datasets}}
\tablehead{
     & \multicolumn{2}{c}{Lab} & \multicolumn{2}{c}{Simulated} & \multicolumn{2}{c}{Field} \\
     & DHM & FLFM & DHM & FLFM & DHM & FLFM
}
\startdata
Specifications & \multicolumn{6}{c}{2048 $\times$ 2048 pixels at 15 frames per second}\\
Observations & 41 & 15 & 360 & n/a & \multicolumn{2}{c}{$\sim$137\tablenotemark{a}}\\
Total frames & 12,300 & 4,500 & 108,000 & n/a & \multicolumn{2}{c}{241,200}\\
Total Size (GB) & \ed{50.4} & \ed{18.5} & \ed{433.0} & n/a & \multicolumn{2}{c}{\ed{1980.0}}\\
Labeled Tracks & 778 & 199 & 15712\tablenotemark{b} & n/a & \multicolumn{2}{c}{n/a}\\
Labeled Motile Tracks & 213 & 62 & 7815\tablenotemark{b} & n/a & \multicolumn{2}{c}{n/a}\\
Purpose & \multicolumn{2}{c}{Quantitative eval.} & \multicolumn{2}{c}{Sensitivity study} & \multicolumn{2}{c}{Field eval.}\\
\enddata
\tablenotetext{a}{Longer observations split into 300 frame observations.}
\tablenotetext{b}{Labeled by nature of simulation.}
\end{deluxetable}


\subsection{Lab Data} \label{sec:obs_data}

The Lab dataset consists of 41 \acrshort{dhm} and 15 \acrshort{flfm} standardized observations of lab-prepared samples. Well-behaved observations without instrument artifacts were chosen in order to quantify our system's performance on realistic data meeting the assumptions described in \ref{sec:osia_requirements}, representing nominal mission operations. Samples included varying densities of \textit{\acrlong{bsub}}, \textit{Chlamydomonas}, \textit{Euglena gracilis}, \textit{Shewanella oneidensis}, and unknown organisms in water samples returned from the field. For the \acrshort{flfm}, fluorescence was induced in select samples via fluorescent stains (e.g., Syto-9 for nucleic acids or FM1-43 for cell membranes), while in others, autofluorescent organisms (e.g., chlorophyll containing algae) were innately visible. Observations were taken with sample chamber flow rates ranging from \SI{0}{\micro\liter\per\second} to \SI{5}{\micro\liter\per\second} to evaluate performance over a diversity of sample processing approaches.

To evaluate the performance of the particle tracker and motility classifier (discussed in Sections \ref{sec:method_tracking} and \ref{sec:method_biosig}), salient particles were manually tracked throughout each observation and annotated as motile or non-motile. Labels were generated by external labelers from Labelbox, a data annotation company. To ensure annotation consistency and quality, we provided the labelers with a labeling guide document and video with a specific annotation protocol. All labels were then reviewed for quality by our research team. In total, 778 and 199 tracks were labeled in \acrshort{dhm} and \acrshort{flfm} data, respectively. All labeled data including raw observations, labeled tracks, and the labeling guide are published online at \dataset[doi:10.48577/jpl.2KTVW5]{\doi{10.48577/jpl.2KTVW5}} \citep{jpl.2KTVW5_2022}.


\subsection{Simulated Data} \label{sec:sim_data}

\ed{To further} characterize the sensitivity of the particle tracking algorithm, we also generated a Simulated \acrshort{dhm} dataset with a wide range of particle densities, \acrfullpl{snr}, and motility characteristics both satisfying and violating the \ed{observation assumptions} in Section \ref{sec:osia_requirements}. This allowed us to quantify how tracker performance degraded as particles became more crowded and the \acrshort{snr} decreased (see Section \ref{sec:results_tracking}). These tracking results also generally apply to the \acrshort{flfm} data, as it uses the same tracking algorithm on an easier task due to the near-zero background fluorescence signal.

\begin{figure*}
    \centering
    \begin{interactive}{animation}{simulation_snr2_20_50_100_particles_stacked.mp4}
    \includegraphics[width=0.8\textwidth]{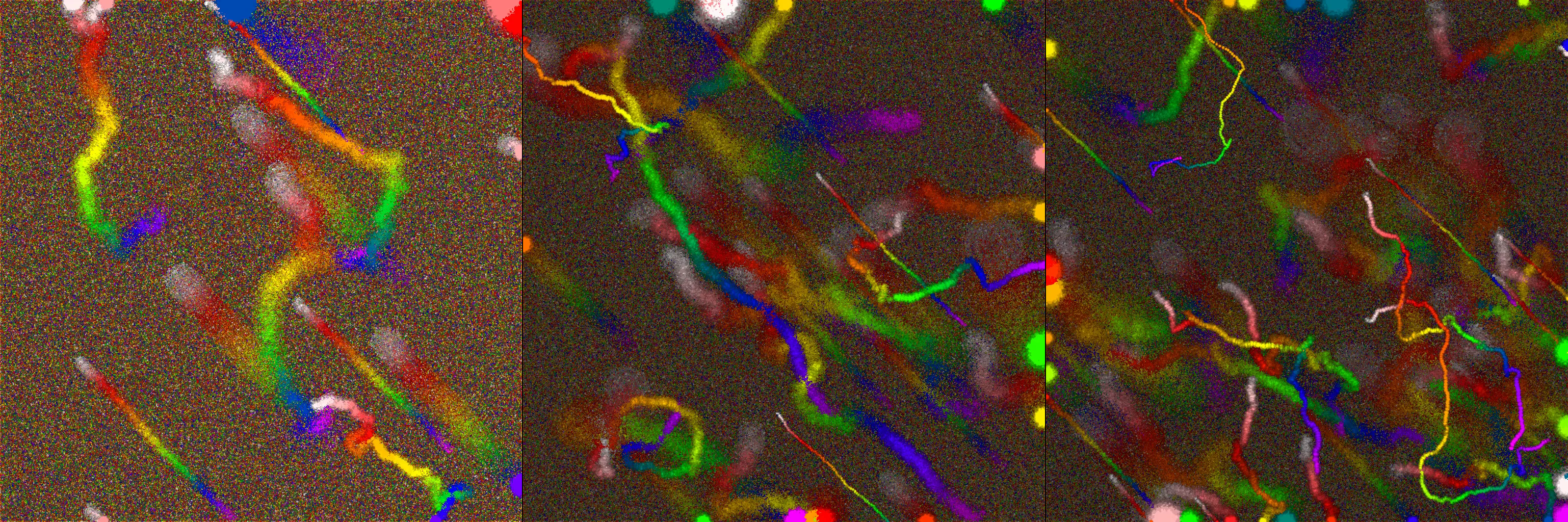}
    \end{interactive}
    \caption{Simulated \acrshort{dhm} data for three separate particle densities (from left to right: low, medium, high) and \acrshort{snr}=2. Increased particle density creates more particle intersections and \ed{makes} tracking of individual particles more difficult. \ed{This figure is available as an animation when the paper is viewed online and shows the \acrshortpl{mhi} followed by 20 seconds of raw simulated data used to generate each \acrshort{mhi}.}}
    \label{fig:simulated_data_example}
\end{figure*}

\acrshort{dhm} observations were simulated by first generating particle tracks, then rendering synthetic particles in individual frames to produce an observation (see Figure \ref{fig:simulated_data_example} for examples). Non-motile tracks were generated assuming simple Brownian random motion, while motile movement tracks were generated with a \acrfull{var} model \ed{fit to} labeled tracks from \textit{Chlamydomonas} observations. A movement bias was then added to both non-motile and motile tracks to simulate smooth flow in the sample chamber. Each particle was then rendered along the specified tracks as an Airy pattern with a fixed (randomly selected) size and brightness. This resulted in observations with a variety of particle densities and \acrshort{snr}. With this simulator, 20 observations were generated for each combination of three particle densities and six \acrshortpl{snr}, for a total of 360 observations. This simulation procedure can be reproduced using the code and \acrshort{var} models in our GitHub repository \citep{OWLS-Autonomy-GitHub}.

\subsection{Field Data}\label{sec:field_data}

The \acrshort{owls} team conducted a week-long field test of the integrated science instruments, \acrshort{osia}, and compute hardware at Mono Lake, CA. Mono Lake is a common analog site for ocean worlds, notable for its high salinity as would be anticipated in samples from Enceladus or Europa \citep{mora2022detection, ferreira2018analysis}. Our team's primary objective was to characterize \acrshort{osia} performance in a field setting and determine where future development is needed for mission infusion. \ed{These results are described in Section \ref{sec:results_field}}. Our secondary objective was to use the generated \acrshortpl{asdp} to assist the science and instrument teams to quickly identify and analyze any biosignatures in collected lake water. Each sampling day consisted of collecting water samples early in the morning from Mono Lake's Station 6 \citep{humayoun2003depth}, recording raw data with the six \acrshort{owls} instruments throughout the day, and applying \acrshort{osia} to analyze recordings in the evening. While scientists and instrument operators had immediate access to each observations, the sheer volume of data collected meant the \acrshort{osia} system remained the fastest method to detect biosignatures, identify any instrument issues, and generate a report to facilitate planning for the next day.

\ed{At the field side}, the particle density of the water samples was extremely high. This led to many particle overlaps within the \acrshort{dhm} data. However, the proportion of those particles exhibiting autofluorescence varied widely. Water collected at 5m contained many autofluorescent particles, while water collected at 35m was approximately 23 times less dense. Presumably, this difference was due to fewer photosynthetic organisms in the deeper (low-light) conditions. After repeatedly detecting high particle densities for the first three days, the science team prepared a diluted sample on the last day of field work. Unfortunately, an instrument issue prevented that data from being recorded properly. Section \ref{sec:results_field} describes results from the field test in detail.

%% file: sections/5_results.tex



\section{Results}

We describe the performance of \acrshort{helm} and \acrshort{fame} \ed{with three approaches} to substantiate a \acrfull{trl} of 5, as required to participate in mission proposal inclusion. \ed{First, we describe our results, takeaways, and lessons learned from evaluating the \acrshort{osia} during a field test of the integrated \acrshort{owls} platform at Mono Lake.} \ed{Second, we quantitatively show that we satisfy the field test requirements described in Section \ref{sec:requirements} by evaluating \acrshort{helm} and \acrshort{fame} on the labeled lab dataset described in Section \ref{sec:obs_data}.} Finally, we leverage both our Lab and Simulated datasets described in Section \ref{sec:sim_data} to further characterize the sensitivity of the \acrshort{osia} system and its submodules.

\subsection{Mono Lake Field Test} \label{sec:results_field}

\ed{The \acrshort{owls} team conducted a week-long field test of the integrated instruments, \acrshort{osia}, and compute hardware at the Mono Lake, CA. The purpose of this test was to evaluate the \acrshort{owls} platform at a relevant analog site and expose needed improvements for mission infusion. (See Section \ref{sec:field_data} for details on the data collected). Below, we describe three key lessons learned from the field test.}

\begin{figure}[ht]
    \centering
    \includegraphics[width=0.4\textwidth]{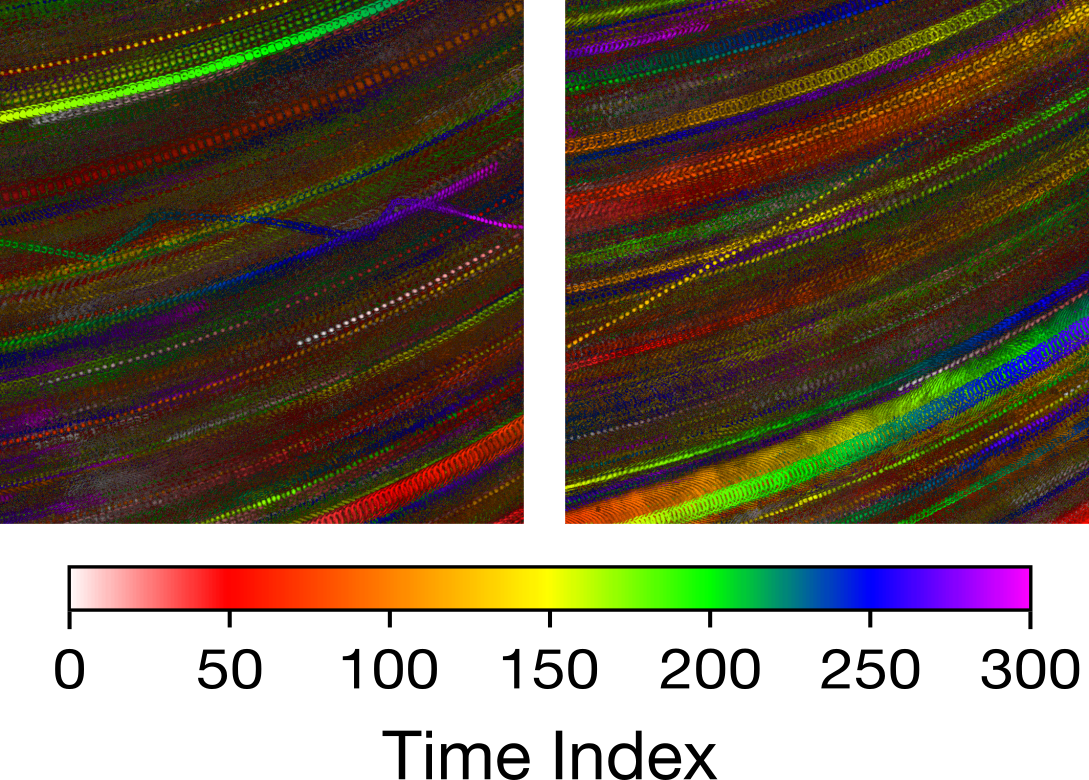}
    \caption{Two motile organisms were captured in \acrshortpl{mhi} during Mono Lake testing. The left \acrshort{mhi} shows a clear zig-zagging movement pattern starting around $t=200$. The right \acrshort{mhi} shows a cell swimming in a consistently different direction than the passively drifting background particles starting around $t=100$.}
    \label{fig:field_mhi_dhm}
\end{figure}

\begin{figure}[ht]
    \centering
    \includegraphics[width=0.4\textwidth]{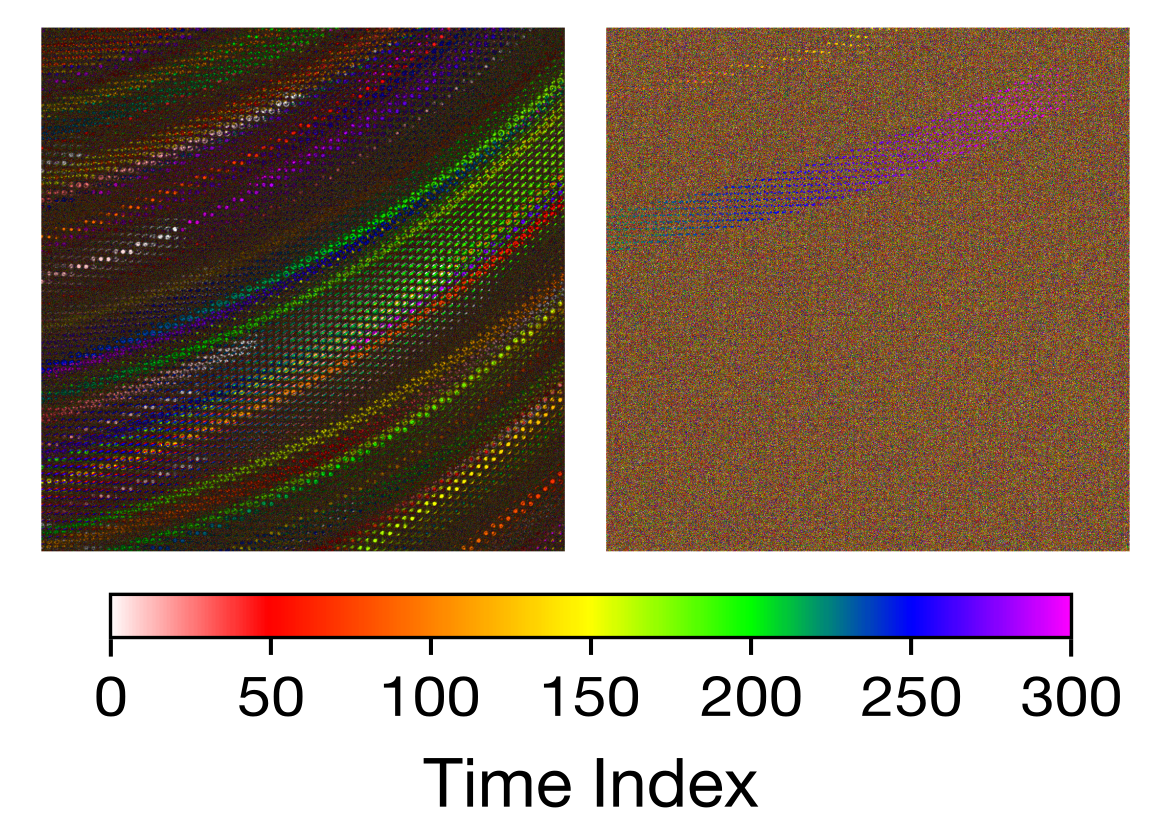}
    \caption{Autofluorescent cells captured at depths of \SI{5}{\meter} (left) and \SI{35}{\meter} (right) in \acrshortpl{mhi} at Mono Lake testing. The \acrshort{osia} detected approximately 23x more autofluorescent particles in the 5m sample (where light is more abundant). The large majority of fluorescent particles had spectral signatures consistent with chlorophyll. No motile autofluorescent microorganisms were detected at Mono Lake during the field trial.}
    \label{fig:field_mhi_vfi}
\end{figure}

First, \acrshort{osia} expedited scientific discovery especially when biosignatures were rare. Motility, for example, was exceptionally sparse in the recorded \acrshort{dhm} data. Only two unambiguously motile organisms were observed, appearing for \SI{12}{\second} in several hours of recorded data (Figure \ref{fig:field_mhi_dhm}). We were able to identify these two motile organisms within about \SI{10}{\minute} of reviewing the \acrshort{helm} \acrshortpl{asdp} at the field site. An abundance of autofluorescent cells was identified in the \acrshort{flfm} data with two examples shown in Figure \ref{fig:field_mhi_vfi}. Mono Lake is known to have high concentrations of \textit{Picocystis} --- a type of green algae --- so high concentrations of chlorophyll were expected \citep{Phillips_2021}. However, \acrshort{fame} was able to track fluorescent particles and provide a density estimation for the two depths sampled (\SI{5}{\meter} and \SI{35}{\meter}), \ed{demonstrating requirement L4-4 in Table \ref{tab:reqs4}}. Overall, the \acrshort{osia}'s ability to rapidly direct attention to the most scientifically relevant data proved valuable in the field setting. \ed{For future life detection missions, this same capability could} enable an efficient \acrshort{conops} strategy.

Second, the data validation and contextual products (see Section \ref{sec:results_valid}) helped the team efficiently react to the field environment; water samples regularly violated one or more of the data assumptions outlined in Section \ref{sec:osia_requirements}. During the first day at Mono Lake, \acrshort{helm} and \acrshort{fame} identified that the water contained particle densities well above the expected range and caused difficulty in tracking particles. Using this insight, the science team planned and carried out sample dilution on day four of the field campaign, \ed{demonstrating requirement L3-4 in Table \ref{tab:reqs3}}. We expect that these automated checks will prove at least as useful in a mission scenario as they did in the field. They could ensure that low-quality data (potentially with incorrect biosignature assessments) does not squander data bandwidth, and that \textit{in-situ} environmental conditions are efficiently communicated to ground teams to inform decisions about instrument operations and \acrshort{osia} reconfiguration. \ed{While reconfiguration of \acrshort{helm} or \acrshort{fame} was not necessary, the capability to do so by editing a plain-text configuration file was available, demonstrating requirement L3-5 in Table \ref{tab:reqs3}.} Future work on \acrshort{helm} and \acrshort{fame} will explore a wider array of data quality and contextual products to more thoroughly capture real-time data characteristics.

Third, the field campaign reinforced the need for a spectrum of \acrshortpl{asdp} ranging from lightly-processed summary products (e.g. \acrshortpl{mhi}) to thoroughly processed, extracted products (e.g. tracks and particle profiles). As mentioned, the Mono Lake field samples contained high particle concentrations well beyond the system's design parameters, simulating a mission instrument miscalibration event. On one hand, \acrshort{fame} performed well under these conditions as only a \ed{small fraction of} particles possessed strong fluorescent signatures. On the other hand, \acrshort{helm}'s tracking performance suffered due to frequent particle crossings, which hindered motility classification. Still, we were able to leverage manual inspection of the \acrshortpl{mhi} to rapidly identify the two motile organisms present. As future missions will face similarly unpredictable conditions either initially or as instrument performance degrades, an \acrshort{osia} strategy that incorporates a range of data summarization techniques will be required.

\subsection{Observation Summarization} \label{sec:results_summarization}

The ability of \acrshort{helm} and \acrshort{fame} to extract scientific content into \ed{summary} data products with a reduced data volume is their primary means \ed{of alleviating bandwidth constraints for} missions to ocean worlds. To quantify this capability, we ran both systems on the complete lab-observed dataset described in Section \ref{sec:obs_data} and computed the ratio of the data volume of raw observations to the ``downlink ready'' \acrshortpl{asdp}. Table \ref{tab:asdp_size} summarizes averages over these values.

\begin{deluxetable}{lDDDD}
\tablecaption{Overview of \acrshort{helm} and \acrshort{fame} \acrshortpl{asdp} on the Lab dataset. The low-\ed{bandwidth} configuration achieves the best \ed{data reduction} ratio possible, while the high-\ed{bandwidth} configuration includes more \ed{particle} portraits.\label{tab:asdp_size}}
\tablehead{
     & \multicolumn{8}{c}{Average Data Volume} \\
     \colhead{Autonomous Science Data Products}  & \multicolumn{4}{c}{HELM Configurations} & \multicolumn{4}{c}{FAME Configurations}\\
      (ASDP) & \twocolhead{Low (kB)} & \twocolhead{High (kB)} & \twocolhead{Low (kB)} & \twocolhead{High (kB)}
}
\decimals
\startdata
Validation Products                          & 3.7 & 3.7 & 2.4 & 2.4 \\
Motion History Image (MHI)\tablenotemark{a} & 355.4 & 355.4 & 360.6 & 360.6 \\
Particle Tracks                              & 115.5 & 115.5 & 25.9 & 25.9 \\
Particle Portraits\tablenotemark{b}         & 191.8 & 959.0 & 171.0 & 855.0 \\
DD, SUE, DQE                                 & 0.2 & 0.2 & 0.2 & 0.2 \\
Total data volume per sample                 & 666.6 & 1433.8 & 560.1 & 1244.1 \\
\hline
Raw Data & \multicolumn{4}{c}{\ed{1258393.8}} & \multicolumn{4}{c}{\ed{1258393.8}} \\
Lossless Compressed (ZIP)& \multicolumn{4}{c}{\ed{1094722.3}} & \multicolumn{4}{c}{\ed{\hphantom{0}523100.4}} \\
\hline
\ed{Data Reduction} (raw/ASDP) & $\mathbf{1887}.\mathbf{8}\times$ & $877.7\times$ & $\mathbf{2246}.\mathbf{7}\times$ & $1011.5\times$ \\
\ed{Data Reduction (ZIP/ASDP)} & $1642.2\times$ & $763.5\times$ & $933.9\times$ & $420.5\times$ \\
\enddata
\tablecomments{Reported data volumes averaged over the observed lab dataset.}
\tablenotetext{a}{The resolution of the MHI can be configured to accommodate bandwidth limitations.}
\tablenotetext{b}{Low bandwidth config keeps one portrait per track. High bandwidth config keeps five portraits per track.}
\end{deluxetable}

Both \acrshort{helm} and \acrshort{fame} are able to produce \acrshortpl{asdp} three orders of magnitude smaller than the original raw data, achieving a \ed{data reduction} ratio of 1887.8 and 2246.7, respectively. \ed{These} results satisfy the \ed{data reduction} requirement \ed{(Table \ref{tab:reqs3} L3-1)}. \acrshort{fame} achieves a higher average \ed{data reduction} ratio because the \acrshort{flfm} only observes fluorescing particles, which are rarer and therefore generate fewer particle track and portrait products. Table \ref{tab:asdp_size} notes in (a) that the \acrshort{mhi}'s resolution or lossy compression quality could be increased or decreased depending on the needs of the specific use case. To demonstrate this, the number of particle portraits taken per track was adjusted between the \ed{``low-bandwidth'' and ``high-bandwidth''} configurations, as described in note (b). This allows missions the flexibility to make tradeoffs, such as downlinking \ed{fewer observations but with more} particle portraits per observation.

The choice of summarization configuration is made with respect to each mission's global constraints as well as the science team's current needs. For example, a \ed{surface mission on Mars} might initially select more and higher-fidelity particle portraits for motile findings, and hence a lower \ed{data reduction} ratio, given \ed{the} relatively high bandwidth availability and the initial anticipation that high-priority findings will be rare. However, if a site proved rich in motility signatures, more \ed{aggressively} compressed results \ed{(providing} more observations per downlink cycle\ed{)} might be preferred to quickly understand the diversity of a site. Finally, the science team might choose to return a very high fidelity record of a few select observations of most interest, spending their bandwidth to defensibly validate and verify previously summarized findings. This scenario emphasizes the ability to leave raw observations onboard and flexibly reprocess them with differing levels of summarization, responding to an evolving science focus and a growing understanding of the environment both globally and for a local sampling site. 

\begin{figure}[ht]
    \centering
    \begin{interactive}{animation}{videos/2021_02_03_dhm_true_low_chlamy_grayscale_lab_14_visualizer_cropped.mp4}
    \includegraphics[width=0.4\linewidth]{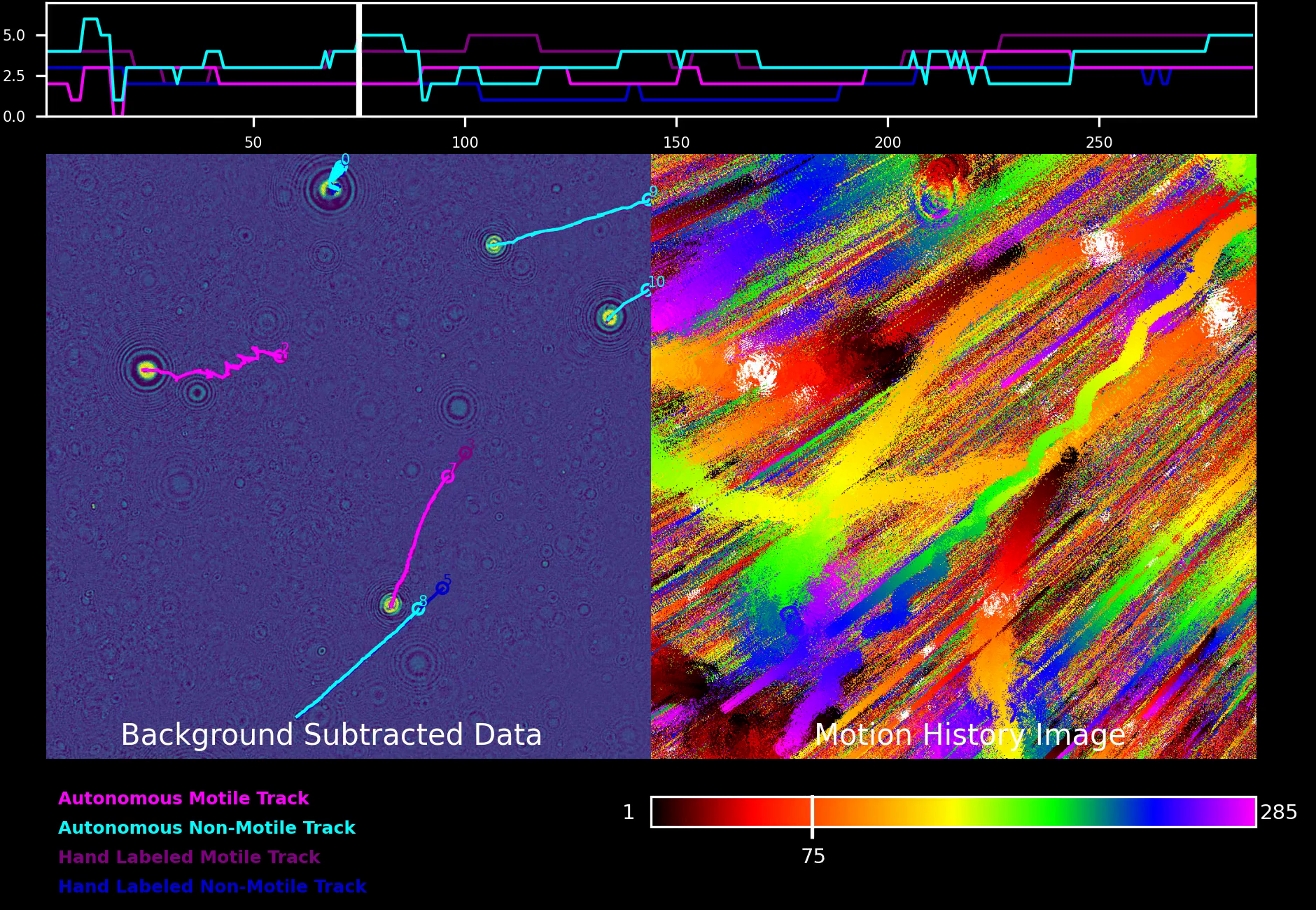}
    \end{interactive}
    \caption{\ed{\acrshort{helm} and \acrshort{fame} can generate a summary animation of particle motion for science teams to visualize any captured biosignatures. \textbf{Left:} Particles are tracked and assessed for motility in a Lab dataset \acrshort{dhm} observation of \textit{Chlamydomonas}. The animation shows non-motile and motile particles being tracked and classified (as bright cyan and magenta lines, respectively) as water flows past the field of view. \textbf{Right:} The \acrshort{mhi} provides a visual summary of the observation. It is similar to \acrshortpl{mhi} shown previously except that the pixels corresponding to the current frame’s max intensity changes are filled in with white. \textbf{Top:} The number of motile and non-motile particles for the current video frame are displayed to provide context. This figure is available as 20-second animation when viewed online.}}
    \label{fig:full_visualization}
\end{figure}

Despite \ed{no bandwidth constraint,} the terrestrial field trial \ed{underscored the need for} rapid understanding and focus of attention \ed{as would be necessary for} planetary use cases. Hundreds of gigabytes of microscopy frames representing hours of \ed{observations were recorded each day.} As instrument scientists and operators were focused on data collection and occasional hardware debugging, they lacked the time to manually and thoroughly review the entire observational record in real-time. \ed{Once generated}, the \acrshortpl{asdp} generated by the \acrshort{osia} \ed{enabled the science team} to review each day's full set of observations in 10-20 minutes. After \ed{identifying scientifically interesting} observations, we could \ed{then} show the science team the corresponding set of video summaries containing the \ed{full} \acrshort{osia} processing results (see Figure \ref{fig:full_visualization} \ed{for a visualization example}). This enabled better planning decisions for the next day's activities and improved the team's understanding of rare events buried within the \ed{vast} observational record.

\subsection{Observation Prioritization} \label{sec:results_prio}

While high data summarization and \ed{reduction} rates allow missions to downlink more observations, content-based awareness offers the ability to \ed{queue observations for downlink in an order that will best satisfy} mission science objectives \ed{or accelerate ground teams' understanding of the environment}. As described in Section \ref{sec:method_prio}, our \acrshort{osia} system produces the \acrfull{sue}, \acrfull{dd}, and \acrfull{dqe} \acrshortpl{asdp}'s to inform \acrshort{jewel}'s prioritization \ed{order} of data products. To demonstrate prioritization, we apply \acrshort{helm} to the Lab \acrshort{dhm} dataset described in Section \ref{sec:obs_data} and prioritize the summarized observations, then analyze the resulting order in terms of successful content recognition.

\subsubsection{The Science Utility Estimate} \label{sec:results_sue}

\begin{figure}[ht]
    \centering
    \includegraphics[width=0.6\textwidth]{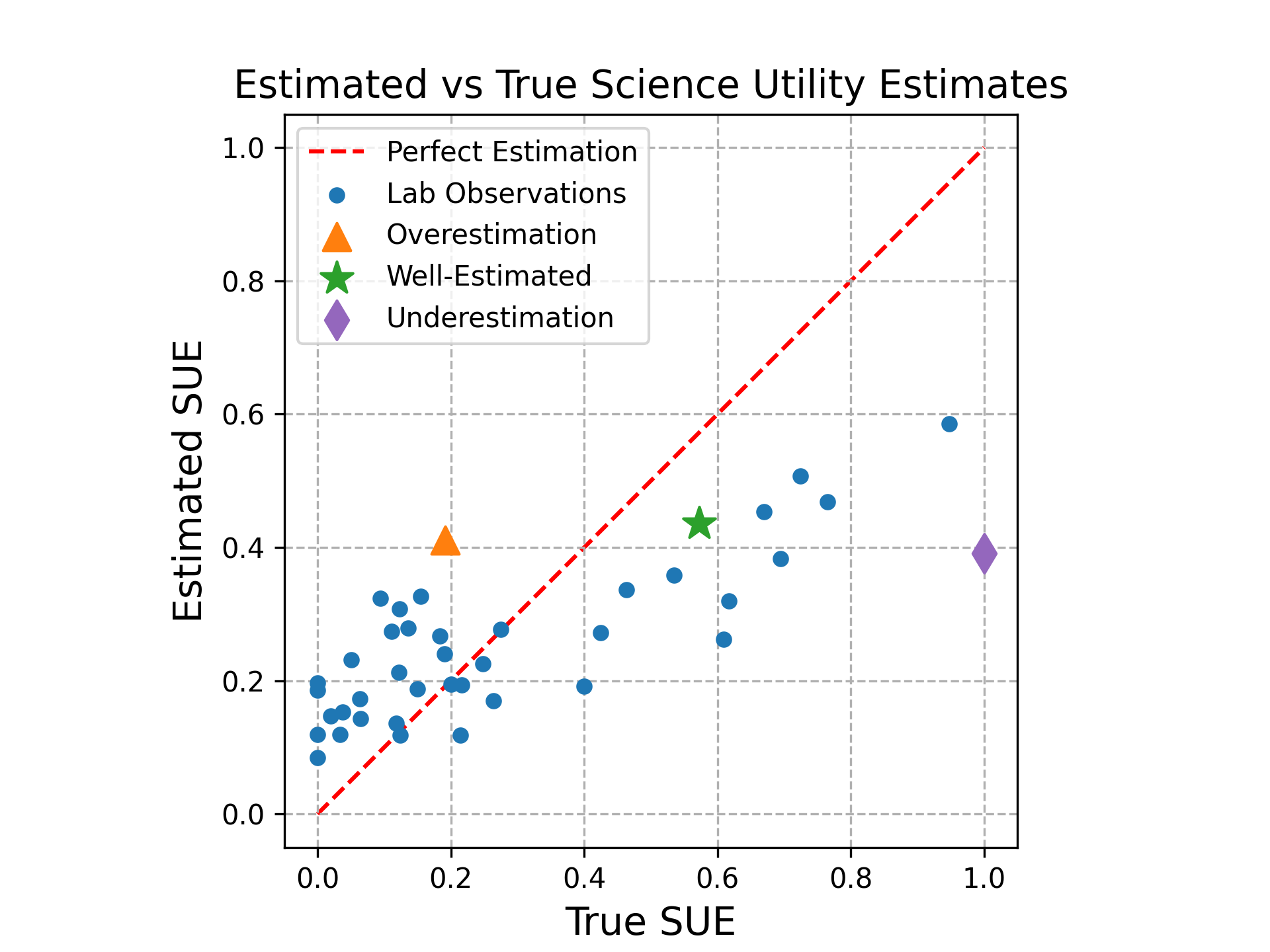}
    \caption{\ed{Comparison of \acrshort{osia}-estimated \acrshort{sue} values to those calculated from ground truth labels on the Lab \acrshort{dhm} dataset}. The estimated \acrshortpl{sue} exhibit an upwards trend with respect to the ``true'' \acrshortpl{sue} calculated from labeled tracks. However, our system slightly overestimates observations with true \acrshort{sue} values $<0.2$ and moderately underestimates observations with true \acrshort{sue} values $>0.4$. Three observations are highlighted (triangle, star, diamond) to facilitate deeper discussion into system imperfections and their driving causes.}
    \label{fig:suetest}
\end{figure}

At the start of a mission, science teams seek observations that most directly satisfy the mission's primary science objectives. For \acrshort{helm} and \acrshort{fame}, the \acrshort{sue}, a quantified proxy for the scientific value of an observation, helps identify observations that contain motility or fluorescence biosignatures. \acrshort{jewel} can be configured to prioritize by simply maximizing the \acrshort{sue} of returned observations. To evaluate the performance of \acrshort{helm} in estimating the \acrshort{sue}, we compare the \acrshort{osia}-estimated \acrshortpl{sue} to the ``true'' \acrshortpl{sue} calculated from human-provided track annotations (Figure \ref{fig:suetest}). As shown, the system \ed{demonstrates the generation of} skillful \acrshortpl{sue} \ed{for use during prioritization (per Req.~L4-5 in Table \ref{tab:reqs4})}.

While the ideal \acrshort{osia} system could perfectly estimate \acrshort{sue} values, the most important outcome is that observations are prioritized given the proper \textit{relative ordering}. Therefore, we compare the sorted lists of observations by their estimated and true \acrshortpl{sue} with Kendall's rank correlation coefficient (or Kendall's $\tau$). $\tau = 1$ indicates that the desired and actual ordering match perfectly, whereas $\tau = 0$ indicates no correlation between the orderings. \acrshort{helm} is able to achieve $\tau = 0.529$ on the entire lab dataset, rejecting $\tau = 0$ with a $p$-value of $1.09 \times 10^{-6}$. Kendall's $\tau$ could be used by future works improving upon this system to benchmark their \acrshort{sue}-based observation prioritization. Additionally, mission concepts could define their \acrshort{osia} prioritization requirements by determining an acceptable $\tau$ for autonomous downlink prioritization through rigorous trade studies.

To explore more deeply the remaining challenges in our system and observations, we investigated three performance cases from Figure \ref{fig:suetest} to understand how faithfully \acrshort{helm}'s \acrshort{sue} calculations represented the data. First, the orange triangle in \ed{Figure} \ref{fig:suetest} \ed{is a representative observation} from a population of low-interest data (true \acrshort{sue} $ 0 \sim 0.2$) that is overestimated by $0.1 \sim 0.2$. In \ed{this observation, only one true motile track was found during labeling, which} lasted through nearly the entire observation. However, due to a high particle density (exceeding assumptions described in Section \ref{sec:osia_requirements}), \acrshort{helm} identified many tracks with $0.4 \sim 0.6$ motility probabilities. This characterizes the system's response to overcrowding: an overestimation of the utility due to motility. While undesirable, fortunately, this inflation is not sufficient to crowd out legitimately high-interest observations.

Second, the green star is an example where \acrshort{helm} generated an appropriate \acrshort{sue} value. Annotators identified ten motile tracks, with the top five tracks ranging from $100 \sim 250$ frames \ed{in duration}. \acrshort{helm} also identified many of these tracks with similar \ed{durations} and motility probabilities ranging from $0.6 \sim 0.8$. \ed{Even in these nominal observations, the estimated \acrshortpl{sue} tend to be lower than the ground truth, with most samples sitting below the red dashed diagonal of perfect estimation. We attribute this to the ability of human annotators to track motile particles for their entire duration with absolute confidence, while the automated tracker generally captures partial tracks due to confounders, even with existing mechanisms that combine track fragments. Should a mission concept require better \acrshort{sue} accuracy (if abundant high-\acrshort{sue} samples are expected, for example), model calibration (discussed in Section \ref{sec:results_class}) and general tracker and instrument improvements could address this underestimation. For our use cases, however, the relative prioritization order was minimally affected as all samples' \acrshortpl{sue} were consistently underestimated.}

Finally, the purple diamond in Figure \ref{fig:suetest} is a high interest example where the \acrshort{sue} was underestimated. Annotators identified ten motile tracks, with the top five tracks lasting all 300 frames. \acrshort{helm} identified its top five tracks with \ed{durations} ranging around 200 frames, with motility probabilities ranging $0.4 \sim 0.6$. However, the particles were at the limits of detection and sometimes overlapped, leading to track fragmentation. The classifier was also less confident with these motility patterns, consisting of slowly curving paths that were not well-represented in the rest of the training data. This supports the need to further expose the system to a robust set of motility styles, both real and simulated, once mission resources become available for a full flight implementation.

\subsubsection{The Diversity Descriptor} \label{sec:results_dd}

We also include the \acrshort{dd} as a component of prioritization to ensure a diverse set of samples comprise the final prioritized list. Whereas the \acrshort{sue} provides a mechanism to exploit an environment by identifying specific biosignatures, the \acrshort{dd} provides a mechanism to explore samples across a range of conditions. It enables the autonomy to distinguish between different categories of observations independent of the \acrshort{sue}. On a real mission, diversity-based sampling would be most relevant once the primary science objectives of a mission were met, or if the initial \acrshort{sue} was a poor fit for the spacecraft's environment; in such situations, ground teams might desire a holistic understanding of the target environment's variability both for exploration and to help identify new biosignatures for \acrshort{sue} reformulation.

To define the \acrshort{dd} vectors, we identified track metrics that reasonably separated the different observations into similar groups. For \acrshort{helm}, we used a \ed{9-dimensional} vector using the $10^{th}$, $50^{th}$, and $90^{th}$ percentile of the size, speed, and end-to-end displacement of tracks within each observation. For \acrshort{fame}, we use the same percentiles for fluorescent intensity, acceleration mean, and step angle mean. \ed{Therefore, the \acrshort{dd} demonstrated the capability to} separate lab and natural samples as well as different flow conditions and organism characteristics \ed{(per Req.~L4-6 in Table \ref{tab:reqs4})}.

\subsubsection{Prioritization Tuning} \label{sec:results_prio_tuning}

\begin{figure}[ht]
    \centering
    \includegraphics[width=1.0\textwidth]{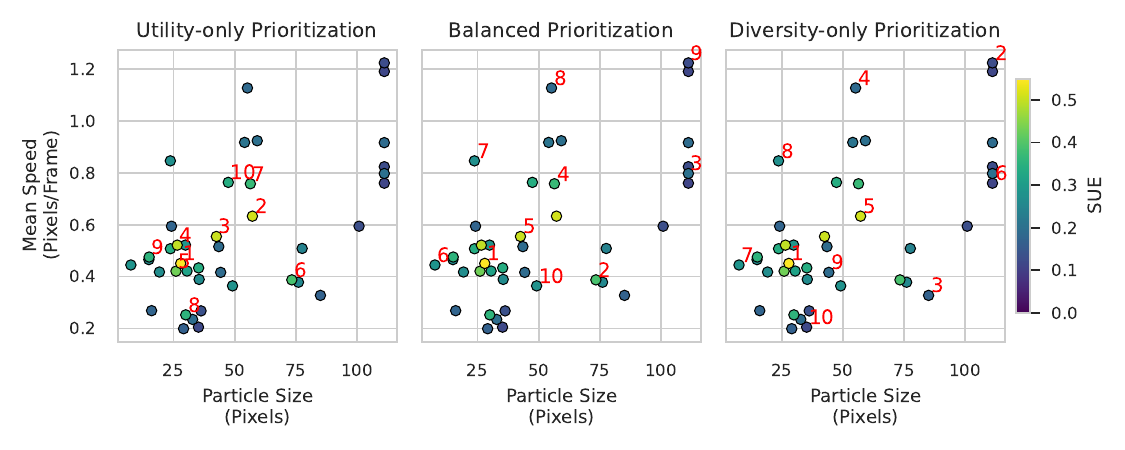}
    \caption{A variety of prioritization strategies are achieved through different weighting of the \acrshort{sue} (indicated by marker color) and the \acrshort{dd} (indicated by marker position). \ed{These plots indicate the top ten prioritized observations (in red text)} from 41 Lab training datasets using three different strategies: scientific utility-only (left), a balance between utility and diversity (middle), and diversity-only (right). The utility-only prioritization selects high \acrshort{sue} experiments without regard for the \acrshort{dd} and chooses several observations judged to contain highly similar in content (as indicated by their proximity in this plot). The middle plot demonstrates a balanced prioritization (with $\alpha = 1.0$ in Eq. \ref{eq:diversity-factor}) where selected experiments have a higher mean \acrshort{sue}, but also provide a sampling of the \acrshort{dd} space. The right plot shows diversity-only prioritization with selections spanning the \acrshort{dd} space without regard to the \acrshort{sue}. Note: the true \acrshort{dd} vector contains 9 elements as specified in Section \ref{sec:method_prio}. We artificially chose two of the the \acrshort{dd} elements here --- the $90^{th}$ percentile of the track particle size and $50^{th}$ percentile track speed --- to visually convey the prioritization concept.}
    \label{fig:prioritization}
\end{figure}

Figure \ref{fig:prioritization} shows three exemplar approaches to prioritizing \acrshort{dhm} observations using different relative weights for the \acrshort{sue} and \acrshort{dd} using the prioritization framework outlined in Section \ref{sec:method_prio}. The color of each point indicates an observation's \acrshort{sue} value, while the axes indicate an observation's relative relationship to others using \acrshort{dd} elements. For intuitive visualization, we have plotted against only two of the original nine \acrshort{dd} elements.

A utility-only prioritization scheme (Figure \ref{fig:prioritization}, left) ranks observations based solely on the \acrshort{sue}. In our evaluation data, this prioritizes lab observations containing dozens of clear, motile \acrshort{bsub} organism tracks. The speed and size of tracked particles in these observations was fairly consistent (comprising most of the grouping toward the lower left). Such a scheme is most beneficial when ground teams are confident in both their restricted interest in specific biosignatures and that the \acrshort{osia} is well tuned to identify these key observations. 

In contrast, a diversity-only prioritization scheme (Figure \ref{fig:prioritization}, right) ranks observations by the difference from the set of all previously transmitted observations (using the \acrshort{dd} hyperspace as a quantitative proxy). After an initial, high-\acrshort{sue} observation is selected, the rest are sequentially chosen by their distance/difference from the previously selected observations. This includes natural samples from Newport Beach, CA on the right side of the plot which had large particles and were relatively fast. A diversity-focused prioritization strategy may be desirable after the primary mission science objectives are achieved or if the initial transmitted observations fail to satisfy a mission's science objectives and contextual information is needed to retune the \acrshort{osia}. They also can provide key awareness of unexpected observation contents critical to inform instrument and \acrshort{osia} reconfiguration.

However, a science team \ed{will often} desire a blend of utility-based and diversity-based prioritization (Figure \ref{fig:prioritization}, middle). Here, the lower left cluster of observations (with many motile organisms) is still sampled, but some observations containing large and/or fast particles (right and top of the plot, respectively) are also included despite lower utility scores. A balanced strategy allows a mission to pursue its stated science objectives while maintaining awareness of unexpected observation content. As the relative weighting between \acrshort{sue} and \acrshort{dd} is a tunable parameter (see Equation \ref{eq:diversity-factor}), ground teams can update the prioritization strategy as their understanding grows. \ed{Taken together, these analyses demonstrate that our \acrshort{osia} system meets the L3-2 requirement for data prioritization, as described in Table \ref{tab:reqs3}}.

\subsection{Data Validation} \label{sec:results_valid}

\begin{figure}[ht]
    \centering
    \includegraphics[width=0.4\textwidth]{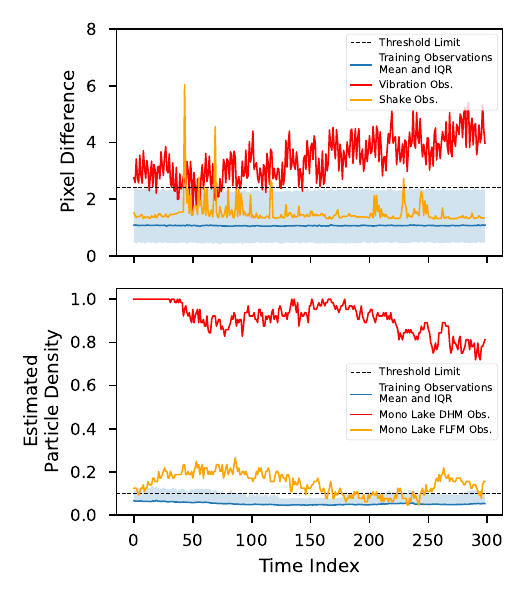}
    \caption{Data validation products help identify quality issues in observations. Here, time traces indicate instrument vibration and shaking (top; red and orange, respectively) as well as excessive particle density in \acrshort{dhm} and \acrshort{flfm} observations from the Mono Lake field campaign (bottom; red and orange, respectively). \ed{The blue trace and shaded region indicates the (expected) mean and interquartile ranges of these same metrics for the training data.} The threshold for each validation check (indicated by the black dotted line) is a tunable parameter that determines if each observation passes or fails. Large frame-to-frame changes and high particle density can both degrade tracking performance and downstream biosignature analysis, so detecting these problems can help influence prioritization and alert ground teams to potential instrument issues.}
    \label{fig:data_validation}
\end{figure}

In addition to science summary products, the data validation step includes simple checks to identify observations that violate expected data characteristics and alert operations teams. If unrecognized, these observations may hamper downstream biosignature analysis. \ed{Refer to Section \ref{sec:method_preproc_valid} and and Table \ref{tab:validation} for the} full list of validation products \ed{generated by \acrshort{helm} and \acrshort{fame}}. Figure \ref{fig:data_validation} illustrates two data quality checks and related observations that violate each. The top of Figure \ref{fig:data_validation} shows \ed{one observation containing} vibration recorded while the \acrshort{dhm} was only partially integrated into \acrshort{owls}. It also includes an observation in which the microscope was physically bumped during recording causing sharp frame-to-frame \ed{shaking. Any microscope movement is important to detect as it can induce fictitious motility-like movement} in observed particles. The bottom of Figure \ref{fig:data_validation} shows a second problem where observations from the field campaign contained high particle densities. In this situation, particles appear to regularly intersect in the raw 2D observations causing degraded tracker performance. \ed{Section} \ref{sec:results_tracking} (and Figure \ref{fig:sim_scores}) provides a more thorough \ed{sensitivity analysis} of how high particle density can degrade tracking performance. For any observation, failing one or more data quality checks translates to a lower \acrshort{dqe} and impacts its assessment during data prioritization. Given similar \acrshort{sue} and \acrshort{dd} values, this ensures high-quality observations are favored for downlink. For each check included in the \acrshort{dqe}, the acceptable bounds and relative weight compared to other checks are configurable to permit retuning by ground teams. \ed{In addition to the \acrshort{dqe}, all validation failures are logged to verbose plaintext reports for operator review. This capability, demonstrated during the field test described in Section \ref{sec:results_field}, satisfies Requirements L4-8,9 in Table \ref{tab:reqs4}.}

\subsection{Computational Performance and Flight Software Integration} \label{sec:results_compute}

HELM and FAME were developed to accommodate limited flight computing resources, including efficient algorithms and flexibility for flight hardware architecture such as multiprocessing acceleration, RAM utilization, and I/O. We anticipate flight computer evolution over the coming years and are baselining platforms similar to the \SI{2.26}{\giga\hertz} Snapdragon 801 SoC with \SI{2}{\giga\byte} of RAM used by the Ingenuity helicopter for near-term flight infusion \citep{balaram2018mars}. \ed{The ``onboard'' computer used during the OWLS field campaign was a \SI{2.1}{\giga\hertz} Intel 3rd generation Core i7-3612QE Quad-Core with \SI{8}{\giga\byte} of RAM.} HELM and FAME were written in Python and integrated into the F-Prime flight software framework \citep{fprime_2021, bocchino2018f} to enable \acrshort{osia} deployment during field testing using the onboard computer.  \acrshort{helm} and \acrshort{fame} contain integration memory and runtime logic to track compute performance. This information is included in instrument suite's telemetry for ground operators and supports runtime optimization on new systems.

To benchmark the \acrshort{osia} capabilities, we used 50 \acrshort{dhm} and 50 \acrshort{flfm} observations recorded during the field campaign (Section \ref{sec:results_field}). These observations, each lasting for 300 frames taken at 15 frames per second, were fully processed using the onboard flight-like computer to determine typical resource usage (as described in \ref{sec:results_field}). Table \ref{tab:benchmarking} provides runtime averages for each stage of the \acrshort{helm} and \acrshort{fame} processing pipelines, in both seconds and as a percentage of the overall runtime. Rapid processing is particularly important in field support use cases, where HELM and FAME are providing focus of attention to field operators. \ed{For such use cases, the team imposed a requirement of processing any field experiment in less than ten times the time it took to record the observation. With field experiments taking 20 seconds to record, this left the HELM and FAME algorithms 200 seconds to process and extract results. Average run times for the HELM and FAME systems were 125.3 and 112.1 seconds respectively, satisfying our requirements (Req.~L4-7 in Table \ref{tab:reqs4}).} RAM utilization is also tracked by the \acrshort{helm} and \acrshort{fame} systems with peak RAM utilization of \SI{1.9}{\giga\byte} and \SI{1.3}{\giga\byte}, respectively, with 4 core multithreading. \ed{To deploy \acrshort{osia} on a mission, specific computational requirements (e.g., on runtime, RAM usage, onboard storage, etc.) should be formulated and tested to ensure that they meet the mission's scientific objectives.}

If needed, future hardware and software upgrades would provide viable pathways to reduce runtime \ed{as components of our image processing pipeline are} amenable to hardware acceleration. For example, pixel-based evaluation steps such as the computationally dominant initial image preprocessing and data validation \ed{could be} offloaded to either GPU or FPGA acceleration \ed{with additional software development}. The \ed{current} \acrshort{helm} and \acrshort{fame} versions also save intermediary data products to the file system between pipeline processing steps to allow processing to be restarted from any pipeline stage in the event of an interruption. However, this is very expensive due to the I/O time required to write $\sim$\SI{5}{\giga\byte} to disk. For missions not anticipating processing interrupts and which possess sufficient RAM, \ed{processing time could be reduced by eliminating these intermediary products}. 

\begin{table}
\centering
\caption{\acrshort{helm} and \acrshort{fame} runtime benchmarks broken down by processing stage. Runtime is provided in seconds and as a percentage of the overall runtime.}\label{tab:benchmarking}
\begin{tabular}{lrccc}
\hline
Processing&\multicolumn{2}{c}{HELM Run Time}&\multicolumn{2}{c}{FAME Run Time}\\
Stage&(seconds)&($\%$)&(seconds)&($\%$)\\
\hline
Preprocessing&41.6&33.2$\%$&38.8&34.6$\%$\\
Validation&62.3&49.7$\%$&62.0&55.3$\%$\\
Tracking&11.8&9.4$\%$&9.7&8.7$\%$\\
Featurization&0.3&0.2$\%$&0.1&0.1$\%$\\
Prediction&0.2&0.2$\%$&0.1&0.1$\%$\\
ASDP Generation&9.1&7.3$\%$&1.4&1.2$\%$\\
\hline
Total Run Time&125.3&100$\%$&112.1&100$\%$\\
Raw Data Collection Time&20&&20&\\
\hline
\end{tabular}
\end{table}

\subsection{Sub-Component Evaluation}

The individual sub-components responsible for particle tracking and motility classification are key to both the summarization and prioritization capabilities discussed above. Below, we present the validation and characterization for these sub-components individually.

\subsubsection{Tracking Performance} \label{sec:results_tracking}

\ed{The algorithms used for particle identification and tracking (described in Section \ref{sec:method_tracking}) have several configurable parameters. We used genetic optimization via \acrfull{toga}\footnote{https://github.com/JPLMLIA/TOGA} to optimize the parameters over a subset of our hand-labeled Lab \acrshort{helm} dataset (described in Section \ref{sec:obs_data}). The optimizer sought to maximize the $\alpha$ tracking quality measure, commonly used to measure particle tracking performance \citep{Chenouard2014-ptc}. $\alpha$ is analogous to recall, and $\alpha=1$ represents a perfect match between labeled and predicted tracks. A high $\alpha$ is analogous to meeting the ``Particle Tracking True Positives'' requirement described in Table \ref{tab:reqs4}, but note that the system can be optimized to meet different mission requirements.}

\begin{figure}[ht]
    \centering
    \includegraphics[width=0.4\linewidth]{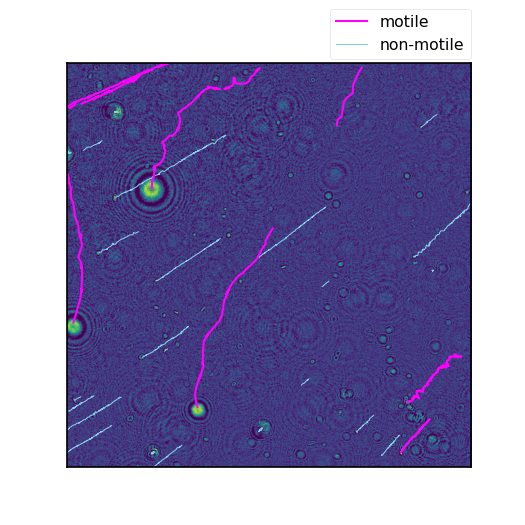}
    \caption{The tracker identifies particles and links them through time to form tracks. Here, tracks are overlaid on the 80th frame of a \acrshort{dhm} observation. \acrshort{ml}-derived motility classifications are indicated by track color. Tracks classified as motile are magenta while tracks classified as non-motile are blue.}
    \label{fig:example_tracks}
\end{figure}

To report the achieved performance of particle track detection, we again use $\alpha$, corresponding to recall, and $\beta$, corresponding to precision (capped at $\beta=\alpha$). On the entire Lab \acrshort{helm} dataset, the optimized tracker reached macro-averages of $\alpha=0.572$ and $\beta=0.474$. We specified $\epsilon = 25$ pixels for these measurements; while prior work has set this value to the Rayleigh criterion \citep{Chenouard2014-ptc} (for our instrument, about $4$ pixels), a larger value is reasonable here as we are processing unreconstructed frames from the \acrshort{dhm} and \acrshort{flfm}. Instead, we defined our value based on our \ed{assumptions} described in Section \ref{sec:osia_requirements}, which bounds the expected particle sizes and the distances between them. \ed{With respect to tracking, we demonstrate} a macro-average true track coverage of $84.7\%$, and $8.45$ false track points per uncrowded observation frame \ed{meeting requirements L4-1,2 in Table \ref{tab:reqs4}}. Figure \ref{fig:example_tracks} shows example tracker output overlaid on a frame of \ed{an} observation. 

\begin{figure}[ht]
    \centering
    \includegraphics[width=\textwidth]{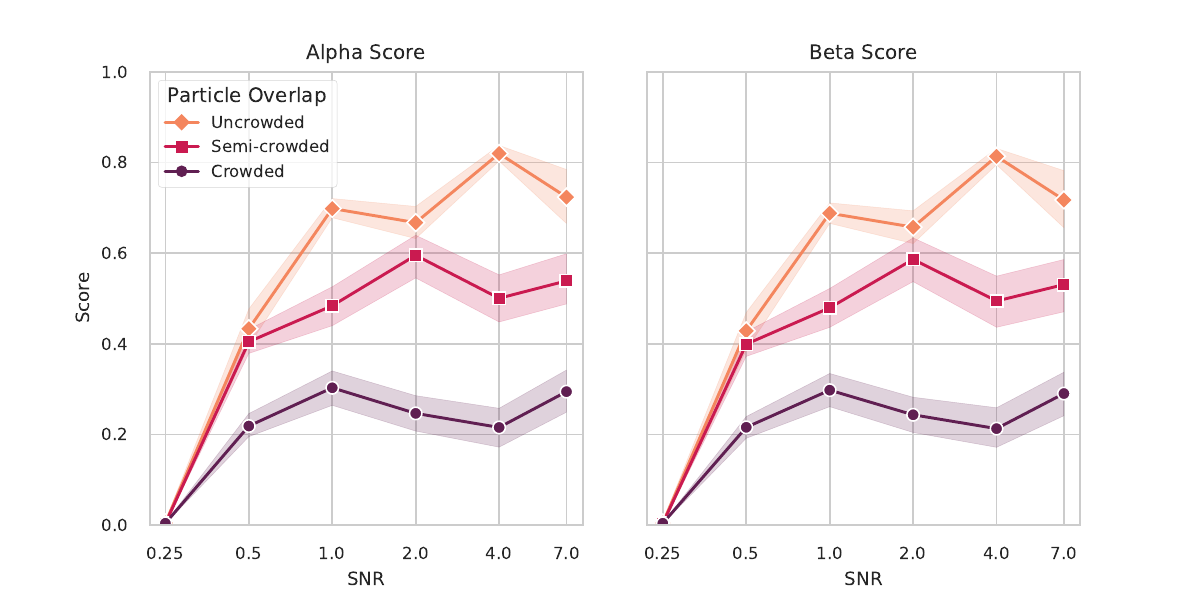}
    \caption{Tracking performance increases with \acrshort{snr} and decreases with particle density. However, changes are limited above \acrshort{snr}=0.5 for the crowded observations, consistently resulting in a $>50\%$ reduction in tracking performance. Note that the maximum value of the beta metric is the alpha metric.}
    \label{fig:sim_scores}
\end{figure}

To assess how tracking performance was affected by particle density and \acrshort{snr}, we generated 360 simulated studies spanning three particle density levels and six \acrshort{snr} levels (Figure \ref{fig:sim_scores}). Higher amounts of particle overlap (e.g., resulting from crowded scenes) led to substantial degradation in tracker performance. This indicates that tracking is limited in observations with many particles even if those particles are visually obvious against the background. In general, once \acrshort{snr} crosses above approximately 0.5, it has a relatively small effect on tracker performance in crowded scenes. For uncrowded and semi-crowded scenes, increasing \acrshort{snr} confers a more steady increase in performance. Experimentally, these results indicate that the capability to dilute water samples (to reduce particle density) is important to ensure optimal particle tracking.

\subsubsection{Motility Classifier Performance} \label{sec:results_class}

We evaluated three \acrshort{ml} architectures for classifying track motility: \acrfullpl{gbt} \citep{friedman2001greedy}, \acrfullpl{rf} \citep{breiman2001random}, and \acrfullpl{svc} \citep{cortes1995support}. These methods were chosen for their simplicity, interpretability, and the computational efficiency of their Scikit-learn implementations \citep{scikit-learn}. We used 5-fold stratified cross-validation to iterate over different combinations of the labeled data (Section \ref{sec:obs_data}) while avoiding data leakage from highly similar experiments. We also applied Bayesian optimization to optimize each method across its hyperparameter search space.\footnote{Hyperparameter ranges for the \acrshort{gbt} models: N. Estimators=[10, 1000]; Max Depth=[1, 20]; Learning Rate=[0.001, 1]. For \acrshort{rf} models: N. Estimators=[10, 1000]; Max Depth=[1, 100]; Class Weight=balanced. For \acrshort{svc} models: C=[1e-4, 1e2]; gamma=[1e-5, 1e3]; Kernel=[linear, RBF]; Class Weight: balanced.} Each model was optimized to maximize the \acrfull{aucroc}. \acrshort{aucroc} was chosen for its ability to represent the precision-recall trade-off; with a model that performs reasonably over a range of posterior probability thresholds, ground teams can make an informed trade-off between conserving data bandwidth and missing a low-confidence motile organism. For a specific flight mission implementation, this tuning would be \ed{thoroughly} re-examined.

\begin{figure*}[ht]
    \centering
    \includegraphics[width=0.4\textwidth]{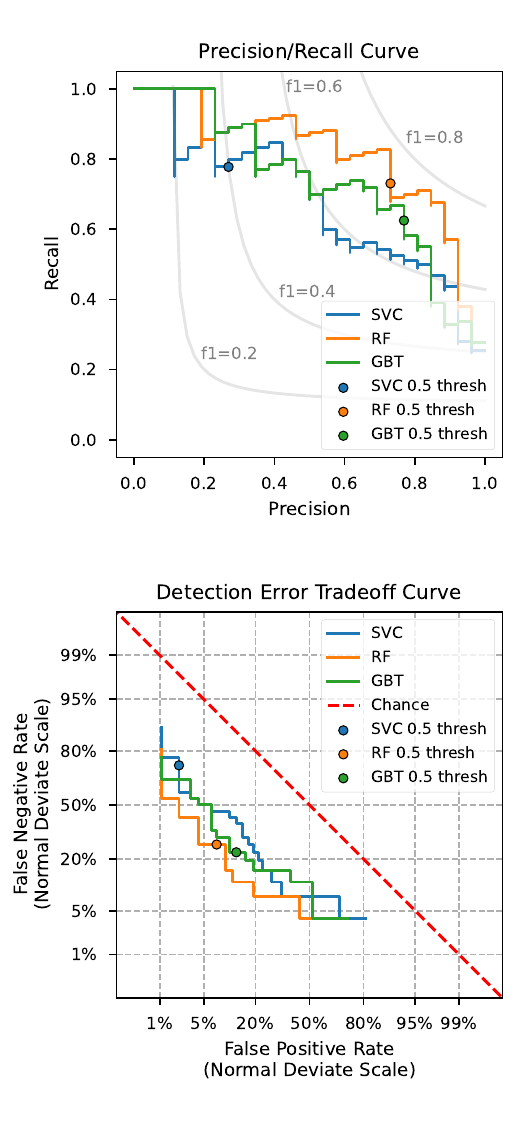}
    \caption{We evaluated the performance of three classical \acrshort{ml} models for identifying motility on a held-out test set: \acrshort{svc}, \acrshort{rf}, and \acrshort{gbt}. \textbf{Top:} Precision-recall curve displays the balance between these two metrics. Here, perfect performance corresponds to the top right corner. The circular marker on each classifier trace corresponds to a 50\% confidence threshold. \textbf{Left:} \acrshort{det} curve displays the balance between false negatives and false positives when traversing a range of confidence thresholds. Perfect performance corresponds to the lower left corner.}
    \label{fig:clf_perf}
\end{figure*}

Figure \ref{fig:clf_perf} shows the performance of each optimized model architecture using \ed{precision-recall and} \acrfull{det} curves\ed{, which assess binary classification performance over a range of decision thresholds}. Precision-recall curves represent the trade-off between the two metrics \ed{and are} is especially informative if the dataset contains a class imbalance. \acrshort{det} curves provide the same information as an \acrshort{roc} curve, but at a scale that better highlights cross-model performance differences and explicitly represents the trade-off between false positives and false negatives for a range of probability thresholds \citep{martin1997det}. In both representations of model performance, the \acrshort{rf} and \acrshort{gbt} models surpass the \acrshort{svc} model. \ed{The \acrshort{gbt}, \acrshort{rf}, and \acrshort{svc} models achieved \acrshortpl{aucroc} of 0.88, 0.92, and 0.86, respectively. \acrshort{rf} reaches the higher overall $F_1$-score (the arithmetic mean between precision and recall) of 0.78, compared to 0.71 and 0.63 for the \acrshort{gbt} and \acrshort{svc}, respectively. A different $F_\beta$ score may be used to evaluate different precision/recall trade-offs as desired.} Elapsed \acrshort{ml} prediction times for all models were well below 1 second, but the \acrshort{gbt} and \acrshort{svc} architectures were approximately $75\times$ faster than the \acrshort{rf} (see Table \ref{tab:benchmarking} for onboard runtimes). Due to the classifier performance and \ed{acceptably} short runtime, we selected the \acrshort{rf} model for further analysis.

\begin{figure}[ht]
    \centering
    \includegraphics[width=0.4\linewidth]{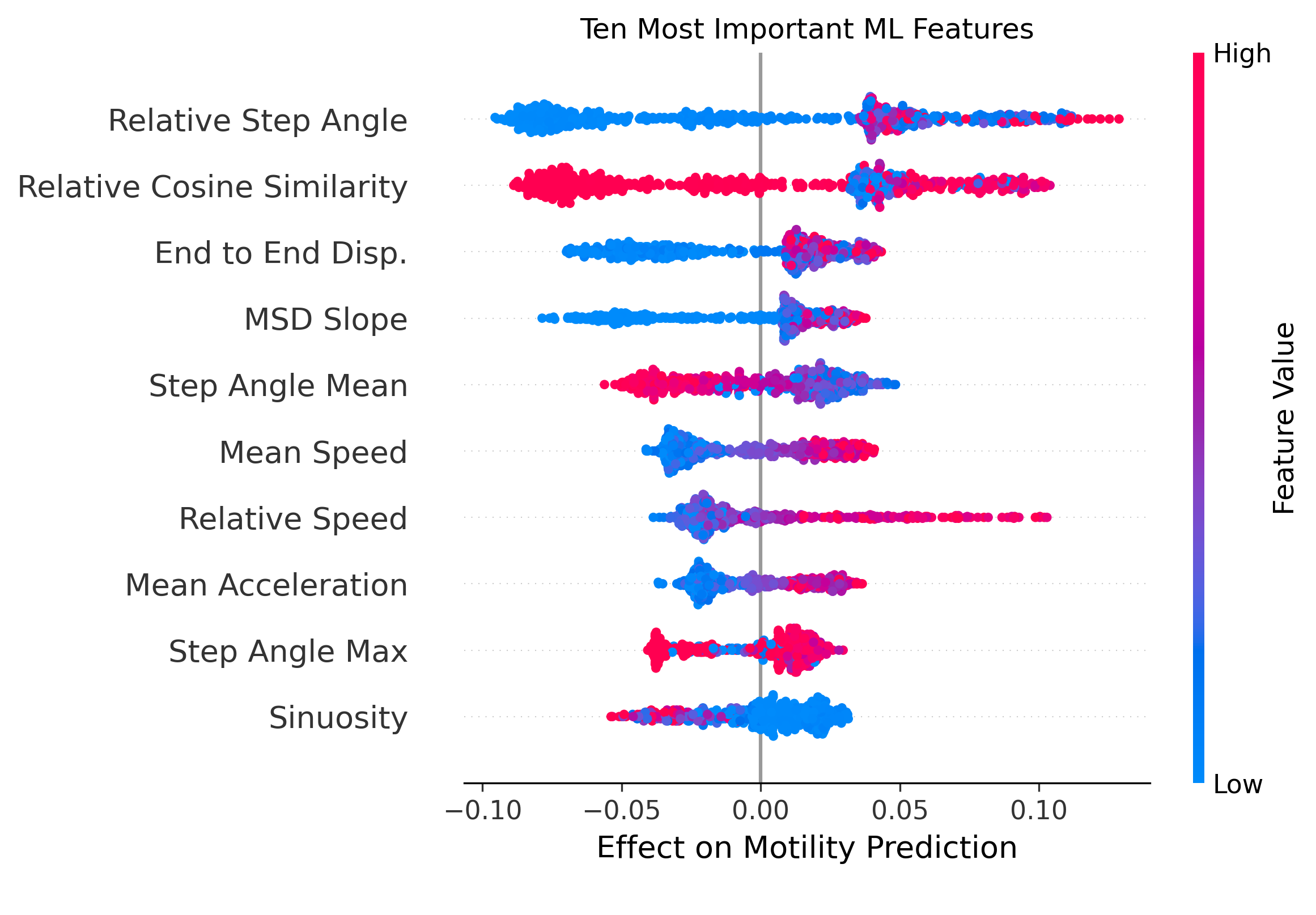}
    \caption{Track features contribute to the motility prediction in different ways, but relative and directional features were frequently ranked among the most important. Using \acrshort{shap}, this plot illustrates how the ten most impactful features (shown in descending order) contribute towards the \ed{\acrshort{rf} model's final motility predictions}. For \ed{each feature row}, the corresponding impact of each feature on the final decision is shown for all tracks (each plotted as one point per row). Note that more important features tend to have wider distributions (along the horizontal axis) corresponding to larger effects on the output probabilities. The color of the points indicates whether a specific track's feature was relatively high or low compared to all others. Taken together, this visualizes how track features contributed to the output classification decisions.}
    \label{fig:clf_features}
\end{figure}

To interpret how individual features (track extracted variables) contribute to classifier decisions, we applied the machine learning explainability method \acrfull{shap} to our trained \acrshort{rf} model. \acrshort{shap} uses a game-theoretic approach to estimate the role of each feature in individual predictions \citep{lundberg_2017shap}. Figure \ref{fig:clf_features} shows the \ed{\acrshort{shap}} values for the ten most impactful features used by the \acrshort{rf} model. Here, negative \acrshort{shap} values (left of center) indicate data points where the feature value corresponded to a push toward non-motile predictions, while positive values (right of center) indicate tracks where the feature value corresponded to a push toward a motile prediction. For an intuitive example, \ed{tracks with a} low mean speed or relative speed -- indicating tracks that moved slower overall or relative to other tracks in the same observation -- translated to lower motility probability as indicated by the concentration of blue points to the left of center. 

In general, the classifier used a variety of different feature types to make decisions. Relative features, which quantify how similar or dissimilar a track was compared to the population of other tracks in the observation, make up the top two features. This highlights the importance of comparing each track against other tracks observed at the same time. High values for relative step angle, which implies a track changed direction much more than other observed tracks, corresponded to a strong push toward motile predictions. For relative cosine similarity, high feature values --- implying the end-to-end track direction was nearly identical with other tracks --- tended to correspond to non-motile predictions. The importance of relative features in classification decisions also implies the existing system is likely to perform worse for (1) extremely sparse observations where only a single track is observed or (2) cases where multiple motile cells move in unison (similar to a school of fish). Interestingly, ``standard'' metrics related to speed and acceleration (see \ref{sec:method_feature}) were of limited importance; only two of the ten calculated ranked among the top ten. \ed{Others have noted that this} line of \ed{explainability} analysis, seeking to characterize precisely how the model makes its determinations, is a vital step towards gaining mission inclusion because it helps build trust with all stakeholders through familiarity and intuition \citep{Slingerland2022}.

\begin{figure}[ht]
    \centering    
    \includegraphics[width=0.4\linewidth]{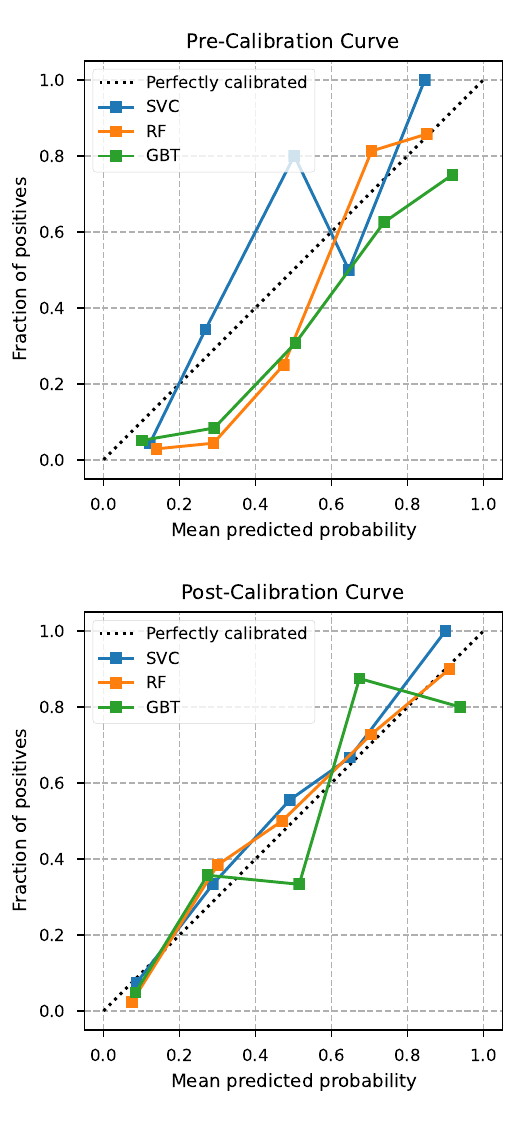}
    \caption{Model calibration provides a mechanism to correct bias in our trained motility classifiers and normalize to a true empirical likelihood estimate. \textbf{Top:} Uncalibrated \acrshort{gbt}, \acrshort{rf}, and \acrshort{svc} classifiers generate predicted probabilities that do not always align with the true class on a held-out test set. Points above the diagonal line indicate low confidence, while points below the diagonal line indicate overconfidence. \textbf{Bottom:} After cross-validated, isotonic calibration the model outputs more reliably represent the true probabilities within each bin.}
    \label{fig:calibration_curve}
\end{figure}

Section \ref{sec:method_biosig} described how the classifier's posterior probabilities may be used for downlink prioritization. \ed{To validate the model outputs, we conducted a calibration assessment by binning posterior probabilities from the test set and checking if the percentage of truly motile tracks aligned with the model's predictions. Since each} model appears under- or over-confident for certain portions of the predicted probability, we applied model calibration to adjust their output probabilities. Figure \ref{fig:calibration_curve} shows the original (uncalibrated) and calibrated \acrshort{gbt}, \acrshort{rf}, and \acrshort{svc} models on the held-out test set. Here, cross-validated isotonic calibration was applied, but other methods also exist. While the application of calibration improves model outputs to more closely match the diagonal line (corresponding to ideal performance), some under- and over-confidence remains. Regardless of whether or not the calibrated model is used in a mission scenario, this procedure improves transparency and interoperability by exposing how faithfully an \acrshort{ml} model's probability estimates align with reality based on the training/validation data available to the system. Note that calibration functions are strictly monotonic, and therefore do not impact model performance metrics shown in Figure \ref{fig:clf_perf}. \ed{Taken together, these results demonstrate that the motility classifier meets requirement L4-3 in Table \ref{tab:reqs4}, with both satisfactory classification performance and estimation of probability of life-like motility.}

\subsubsection{Particle Portraits} \label{sec:results_portraits}

\begin{figure}[ht]
    \centering
    \begin{interactive}{animation}{videos/dhm_flfm_recon_stacked.mp4}
    \includegraphics[width=0.4\textwidth]{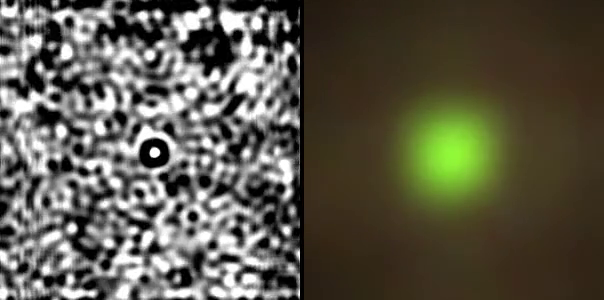}
    \end{interactive}
    \caption{Image reconstructions allow detailed analysis of individual microorganisms. \ed{The reconstructions here were generated using particle portraits extracted from tracks identified at Mono Lake and show \acrshort{dhm} (left) and \acrshort{flfm} (right) images refocused to the depth of the contained particle. The small, central white dot in the left image shows the microorganism responsible for the zig-zagging movement pattern in Figure \ref{fig:field_mhi_dhm} while the right image shows one of the autofluorescent particles belonging to the rare particle class hypothesized to contain green fluorescent protein. This figure is available as a 3-second animation online, which steps through the full z-stack of focal reconstructions to convey the 3D structure of each particle.}}
    \label{fig:example_reconstructions}
\end{figure}

While raw \acrshort{dhm} and \acrshort{flfm} observations are much too large to transmit from ocean worlds, \acrshort{helm} and \acrshort{fame} crop small, full-resolution particle ``portraits'' to extract the raw image data \ed{containing} tracked particles \ed{(as discussed in Section \ref{sec:method_portrait})}. Scientists can reconstruct these portraits to different z-planes and analyze morphology or search for sub-cellular structures in any suspected microorganisms. Since the reconstruction algorithm is computationally expensive (see Section \ref{sec:owls_volumetric_considerations}) and the space of possible particle morphologies is difficult to analyze autonomously with \acrshort{osia}, we include these particle portraits in the downlinked \acrshortpl{asdp} for manual investigation by science teams. Figure \ref{fig:example_reconstructions} shows two example reconstructions from Mono Lake data. While the microorganisms identified during the field campaign were small, spherical, and displayed no obvious sub-cellular details, this capability to extract individual particle sizes would provide valuable detail for microorganisms with non-spherical shapes or resolvable sub-cellular structures. 

%% file: sections/6_infusion.tex



\section{Infusion and Adoption}

\ed{Life detection is an important theme for several mission concepts in the most recent Planetary Science Decadal Survey \citep{Decadal_Survey}. To enable unambiguous life detection, these missions should include multi-instrument payloads like \acrshort{owls} that collect observations relevant to multiple orthogonal biosignatures. Many of these instruments produce large data volumes and could benefit from the type of \acrshort{osia} treatment described in this work. In this section, we seek to directly inform mission formulation to encourage early consideration of these onboard science capabilities. Specifically, we will review potential benefits to consider during formulation trade studies, assess the computational feasibility of \acrshort{osia} for three life detection concepts, and identify future \acrshort{conops} tools needed to successfully deploy \acrshort{osia}. Finally, we will discuss lessons learned through this work and highlight terrestrial science applications that could be leveraged to further mature these novel capabilities for space flight infusion.}

\subsection{OSIA for Mission Concepts}

\ed{In Section \ref{sec:managing_bandwidth}, we discussed how \acrshort{osia} can improve a mission's science return through summarization and prioritization, alleviating the bandwidth barrier often facing planetary missions. To illustrate the impacts of this improvement, we reference our \acrshort{osia} implementation on \acrshort{owls} to estimate how \acrshort{osia} could benefit three relevant life-detection mission concepts.}

\subsubsection{OSIA Enabled Capabilities}\label{sec:mission_concept_benefits}

\ed{Onboard science capabilities may be developed to achieve a variety of goals. We will explore six here: addressing data sufficiency, supporting advanced instruments, leveraging inactive periods, shortening reaction times, retargeting/reprocessing observations, and improving communication robustness. While we discuss these at a high level, detailed, formulation trade studies are needed to both identify which (if any) of these advantages warrant \acrshort{osia} consideration and to quantify any expected benefits. We acknowledge that, for each of these capabilities, additional constraints and considerations such as available power, thermal management, consumable resources such as sampling vessels, time to physically collect samples, and competing instrument observation schedules will ultimately influence infusion. Some of these concerns may be addressed by integrating synergistic autonomy elements such as onboard planning and scheduling \citep{m2020-planner-astra2022}}.

\paragraph{Data Sufficiency}

\ed{For any new mission concept, mission architects must show how the volume and type(s) of observational data collected will address the scientific questions at hand. \acrshort{osia} offers an alternative to the standard practice of transmitting all collected raw instrument data. For example, mission teams may wish to acquire more observations than can be downlinked when searching for rare phenomena and then only transmit those deemed scientifically useful. Alternatively, for a fixed bandwidth, missions can transmit summaries of many more observations than would be possible if returning raw data. Here, a simple metric of success could compute the ratio of total time spent returning an observation to the team (processing + transmission) for traditional and \acrshort{osia}-based approaches (as we do later in Section \ref{sec:infusion_concept_formulation})}. 

\paragraph{Advanced Instruments}
\ed{The combination of \acrshort{osia} and next generation processors will enable the consideration of newer, high data volume instruments that are infeasible for today's planetary mission concepts. Such was the core benefit provided by \acrshort{helm} and \acrshort{fame} --- summarization of high-resolution video enables the search for cellular motility even when exploring bandwidth-constrained science targets like Europa and Enceladus. This mismatch between data collection and data downlink rates also limits the use of other advanced instruments such as high-resolution imaging spectrometers or Raman spectrometers. The combination of both \acrshort{osia} and next-generation processors are likely required for deployment of these instruments beyond Earth's orbit. Modern multi-spacecraft concepts (e.g., cubesat constellations or distributed landers) may also face a similar challenge as their collective data volumes can outstrip today's monolithic spacecraft. When considering \acrshort{osia}, an early metric of success might be to simply assess feasibility; for a given communication bandwidth, concept teams could determine if \acrshort{osia} could conceivably summarize these large observations to fit within bandwidth constraints.}

\paragraph{Leveraging Inactive Periods}
\ed{Missions may include periods of inactivity due to communication downtime or limited observation opportunities. These idle periods may be reclaimed for observation processing to summarize and/or prioritize previously acquired observations for upcoming communication windows or other autonomously-triggered tasks. This could substantially increase the number of observations returned by reducing or eliminating the impact of \acrshort{osia} processing on ``active mission time.'' Such was the primary benefit of the \acrshort{aegis} algorithm deployed on the Curiosity and Perserverence rovers (discussed in Section \ref{sec:past_osia}), which selects rocks for ChemCam sampling after a drive but before a ground-in-the-loop command cycle \citep{francis17:aegis}. One useful metric of success here might be the number additional science data returned when comparing a manually commanded spacecraft to one with \acrshort{osia}.}

\paragraph{Mission Reaction Time}
\ed{The time required to transmit, process, and review raw observations during  traditional ``ground-in-the-loop'' cycles limits the agility of mission teams. Current missions do not determine transmission order according to an onboard estimate of scientific utility, raw data requires (sometimes extensive) ground processing before reaching science teams, and spacecraft cannot often react autonomously to scientific data even if other autonomous systems are in place (e.g., for planning and scheduling). Thus, reactions to instrument data or problems can be delayed to the point of irrelevance. \acrshort{osia}'s ability to assess observations onboard means transmitted data can directly inform ground teams about the state of the scientific environment thereby permitting faster decision making. It also and opens the possibility for more missions to make autonomous decisions based on science data. A simple proxy metric for evaluating \acrshort{osia} might be the reduction in time required to alert ground teams to phenomena of interest.}

\paragraph{Soft Retargeting and Reprocessing}
\ed{When data summarization involves excising regions of interest from the original observation (either autonomously as with \acrshort{helm} and \acrshort{fame} or via manual specification), there is the risk of omitting scientifically-interesting content. By storing the raw data onboard and supporting reconfiguration of the \acrshort{osia}, observations may be reprocessed to better extract the desired content without re-acquiring each observation. Similarly, \acrshort{osia} parameters may be updated to improve summary products and optimize science utility on previously processed raw observations. For this benefit, a possible metric of success is the total number of unique observations, processed, and transmitted to ground. With retargeting and reprocessing, repeat observations (e.g., to correct issues with an initial observation) should be reduced, so the spacecraft can spend more resources sampling the environment for new phenomena.}

\paragraph{Communication Robustness}
\ed{To receive instrument data, transmissions from spacecraft beyond Earth's orbit must go through the \acrshort{dsn}. Therefore, a mission's science yield may be affected if those assets do not meet the availability that was envisioned during formulation. For example, increased rationing of \acrshort{dsn} downlink time during upcoming crewed Artemis missions or malfunctions in one of the aging Martian relay orbiters could impact the transmission viability of interplanetary missions. Where appropriate, \acrshort{osia} is worth exploring as a tool to improve communication robustness during such situations; the capability to prioritize transmission of data with the highest estimated science utility or tune \acrshort{osia} configuration to summarize more aggressively may mitigate mission risks. A method to quantify any benefits might involve first collecting (or simulating) a set of representative instrument data. Using this dataset, a mission formulation team could estimate how science return is affected under a set of increasingly strict bandwidth constraints for conventional vs. \acrshort{osia}-based processing approaches.}

\subsubsection{OSIA Feasibility on Flight Processors}\label{sec:osia_flight_feasibility}

\ed{An critical consideration for the infusion of any \acrshort{osia} is compute feasibility. For some mission concepts, advanced compute resources will be required to enable the autonomy algorithms developed in this work. To inform this need, we provide coarse feasibility estimates for \acrshort{osia} on future missions by considering how the runtime of our algorithms will change depending on the compute platform available. We acknowledge that, short of benchmarking a skillfully ported algorithm to a specific flight compute architecture, cross-processor comparisons are difficult to estimate accurately. Therefore, we limit these analyses to an \acrfull{oom} estimate.}

\ed{We broadly divide our compute comparison into current- and next-generation processors. The first category of processors we consider are those with clock speeds in the \SI{100}{\mega\hertz} range; this includes the RAD750 and LEON class of processors. The RAD750 was deployed on \acrshort{msl} (Curiosity) and M2020 (Perseverance) while LEON processors were used on the LICIACube satellite, which observed DART’s asteroid impact, as well as a number of \acrshort{leo} CubeSats. The second category consists of processors with clock speeds in the \SI{1}{\giga\hertz} range, including the Snapdragon 855 and \acrfull{hpsc}. The Ingenuity helicopter uses an earlier Snapdragon model (the 801), but not that the \acrshort{hpsc} has not been deployed as it was in development as of the time of this work. A previous study by the NEAScout mission quantified onboard science processing performance changes across these two processor classes \citep{lightholder2023benchmarking}. A variety of compute operations demonstrated runtime improvements on the order of $10 \times$ to $220 \times$ (with a total improvement of $50 \times$) when moving from MHz- to GHz-scale processors. Using this analysis, we provide tentative predictions about the runtimes of \acrshort{helm} and \acrshort{fame} on future missions.}

\ed{If deployed, we expect that a flight implementation of our \acrshort{osia} algorithms will run at least as fast on a dedicated GHz-scale processor (e.g., the Snapdragon 855) as on the i7-3612QE third-generation quad-core  processor used in our field trial, as benchmarked in Table \ref{tab:benchmarking}. This is a reasonable assumption as: (1) we expect that a compiled flight-ready implementation of our software will run faster than the current implementation in Python (an interpreted language), (2) the Snapdragon 855 has a faster base frequency (2.96 GHz) than the i7 processor (2.1 GHz), (3) the Snapdragon 855 has 8 physical cores compared to the i7 processor's 4 (hyperthreaded to 8), and (4) the Snapdragon 855 has hardware accelerators (including a \acrshort{gpu}) that could further accelerate image processing steps used on \acrshort{helm} and \acrshort{fame}. (The i7's integrated HD4000 \acrshort{gpu} was not used in this work). Therefore, we can conservatively estimate that \acrshort{helm}'s and \acrshort{fame}'s per-observation processing time will remain in the 100s of seconds for GHz-scale compute platforms (as quantified in Table \ref{tab:benchmarking}). Further, we take the most conservative conclusion from \cite{lightholder2023benchmarking} and estimate a $10 \times$ slowdown for MHz-scale compute platforms, estimating a runtime of 1000s of seconds for this category. Beyond processor considerations, \acrshort{helm} and \acrshort{fame} were designed with RAM, storage, and runtime constraints in mind. Software configuration allows mission operators to adjust parallelism to take advantage of platforms with extensive hardware resources (RAM and CPU cores), while still operating on platforms with limited resources. We acknowledge that many other factors affect compute time, and other mission constraints and risk posture will limit choices around flight processors. Regardless, these estimates indicate that \acrshort{osia} algorithms described in this work could be feasibly deployed on either MHz- and GHz-scale compute platforms.}

\begin{deluxetable}{lcc}
\label{tab:flightcompute}
\tablecaption{\ed{Coarsely estimated \acrshort{osia} runtime on flight processors. Considering MHz- and GHz-scale processor performance and our \acrshort{osia} implementation on the field computer (Table \ref{tab:benchmarking}), we estimate \acrshort{helm}'s and \acrshort{fame}'s observation processing time on flight computers to an \acrfull{oom}.}}
\tablehead{
 & \colhead{MHz Scale\tablenotemark{a}} & \colhead{GHz Scale\tablenotemark{b}}
}
\startdata
Clock Speed (MHz) & 100s & 1000s \\
Typical RAM & MBs & GBs \\
Avail. Hardware Accel. & N & Y \\
\hline
\acrshort{oom} Time per Sample& \SI{1e3}{\second} & \SI{1e2}{\second}
\enddata
\tablenotetext{a}{RAD750, LEON-class \citep{berger2001rad750, lightholder2023benchmarking, sjalanderleon4}}
\tablenotetext{b}{Snapdragon 855, HPSC, \citep{balaram2018mars, dunkel2022iss, doyle2013hpsc, powell_2018_hpsc}}
\end{deluxetable}

\subsubsection{Potential Impacts on Mission Concept Formulation}
\label{sec:infusion_concept_formulation}

\begin{deluxetable}{rrrcc}
\label{tab:science_return_speed}
\tablecaption{\ed{Coarse \acrfull{oom} estimates for time-to-ground speedup ratios of a single \acrshort{dhm} observation (defined in Section \ref{sec:data} and Table \ref{tab:asdp_size}) for three mission concepts, two processor categories, and using \acrshort{helm} summarization (\acrshortpl{asdp} produced by the ``high-bandwidth'' configuration) or ZIP compression. While these estimates provide rough intuition for how compute platforms and the inclusion of onboard processing may benefit science return, a thorough mission architecture trade study must consider many other constraints to evaluate the true impact on any mission concept. Estimated \acrshort{osia} runtimes are from Table \ref{tab:flightcompute}.}}
\tablehead{\colhead{Ref. Mission} & \colhead{ZIP product} & \colhead{ASDP} & \colhead{MHz Scale} & \colhead{GHz Scale}\\
\colhead{(Downlink Rate)} & \colhead{Downlink (s)} & \colhead{Downlink (s)} & \colhead{\acrshort{oom} Speedup\tablenotemark{a}} & \colhead{\acrshort{oom} Speedup\tablenotemark{b}}}
\startdata
Enceladus Orbilander (\SI{34}{\kilo\bit\per\second}) & \num{2.6e5} & \num{3.4e2} & $\approx2$ & $\approx3$\\
Europa Lander (\SI{48}{\kilo\bit\per\second}) & \num{1.8e5} & \num{2.4e2} & $\approx2$ & $\approx3$\\
Martian Rover (\SI{1000}{\kilo\bit\per\second}) & \num{8.8e3} & \num{1.2e1} & $\approx1$ & $\approx2$\\
\enddata
\tablenotetext{a}{$\text{round}\left(\log_{10}{(\text{ZIP Downlink} / (\text{\SI{1e3}{\second} Runtime} + \text{ASDP Downlink})})\right)$}
\tablenotetext{b}{$\text{round}\left(\log_{10}{(\text{ZIP Downlink} / (\text{\SI{1e2}{\second} Runtime} + \text{ASDP Downlink})})\right)$}

\end{deluxetable}

\ed{Existing reference mission concepts aimed at detecting life in our solar system vary greatly in terms of their constraints and proposed \acrshort{conops}. Therefore, it is illustrative to explore how \acrshort{osia} might impact each. We refer to \acrshort{osia} capabilities outlined in Section \ref{sec:mission_concept_benefits} by \textit{italicizing relevant descriptions} below. However, we emphasize that complete formulation trade studies are needed to precisely quantify any potential benefits. Such a study must consider other critical choices (e.g., available power, thermal management, consumable resources, scheduling, sample collection, etc.). While a full trade study is out of scope of this work, we highlight how \acrshort{osia} can enhance the architecture or operations of existing mission concepts. To support this analysis, we will refer to Table \ref{tab:science_return_speed}, which calculates the time-to-ground speeds for computing and downlinking \acrshortpl{asdp} compared to the more conventional approach of downlinking raw data after ZIP compression. We also continue our focus on microscopes for biosignature detection, but the benefits of science autonomy for other instruments are discussed elsewhere \citep{francis17:aegis, Mauceri_2022, Theiling2022_science_autonomy}}.

\paragraph{Enceladus Orbilander}

\ed{The Enceladus Orbilander \citep{orbilander_mission_concept} is a concept with potential \acrshort{osia} applications driven by its extreme distance and relatively long duration of 2 years. While at Saturn, the concept estimates a science data downlink rate limited to \SI{34}{\kilo\bit\per\second}. It describes the deployment of a \acrfull{lds} similar to \acrshort{owls}, and includes a microscope as a high-risk, high-reward instrument. Over the course of 2 years of surface operations, the mission concept specifies 29 microscope observations totaling \SI{0.12}{\giga\byte} of data, and would collect only still images to directly look for cells.}

\ed{In this case, potential \acrshort{osia} applications are driven by the 2-year surface phase. This affords considerable onboard processing time, meaning that a less powerful onboard processor may be permissible given ample \textit{inactive periods} to process each observation. As seen in the first row of Table \ref{tab:science_return_speed}, \acrshort{osia} with a MHz-scale processor would still provide a two \acrshort{oom} speedup over downlinking raw compressed products. As samples are collected passively (dependent on fallout and accumulation rates) or actively (dependent on scooping material from the surface), \acrshort{osia} alone would not enable additional sample collection. However, multiple observations per physical sample could be enabled. Each would then be analyzed and down-prioritized to reduce scientifically redundant information to \textit{improve data sufficiency}. Additionally, \acrshort{osia} could better inform follow-up sample collection. For example, the ``\acrshort{lds} Full'' operational mode specifies two separate sample scooping procedures. Expedient, prioritized return of summarized \acrshortpl{asdp} from the initial observations could \textit{inform reactive decisions} about when and where to collect the second scoop. Finally, the three \acrshort{oom} speedup afforded by more powerful onboard compute would enable \textit{advanced instrument observations} such as microscopy videos. This could enable the search for motility biosignatures as demonstrated with \acrshort{helm} and \acrshort{fame}.}

\paragraph{Europa Lander}

\ed{The second reference mission is the Europa Lander mission concept \citep{europa_mission_paper_2022} with \acrshort{osia} applications driven by its substantial distance, short mission duration of 30 days, and a desire for a combination of autonomous operations and decision making by the ground science team. The concept describes a Baseline Model Payload that includes optical and atomic force microscopes and plans for only five total samples. Communication bandwidth is estimated at \SI{48}{\kilo\bit\per\second}, which would be shared between science and telemetry. The authors noted in a previous concept that the communication rate could ``represent the biggest bottleneck in the timely return of decisional data'' \citep{europa_mission_concept_2017}. The latest concept specifically discusses autonomy to address this issue, including ``autonomy and machine-learning techniques...to allow the onboard system to replan communications activity, based on assessment of instrument data and priority measurements'' \citep{europa_mission_paper_2022}.}


\ed{With an expected combined science and engineering data return of $>$ \SI{187.5}{\mega\byte} over the 30-day mission, \textit{data sufficiency} and \textit{communication robustness} are important considerations for this mission concept. Compared to the planned approach of raw-data transmission, \acrshort{osia}'s ability to summarize and prioritize data could improve the chances that any encountered biosignatures are successfully transmitted. As in the second row of Table \ref{tab:science_return_speed}, \acrshort{osia} with a GHz-scale processor would provide a three \acrshort{oom} speedup over downlinking raw compressed products. In addition, the short lifetime of the mission implies that operators will be under significant time pressure for any ground in the loop operations. Therefore, the rapid transmission of any summarized scientific insights is likely to be valuable for \textit{informing and shortening decision-making processes}. The mission concept also states that, due to the constraints of direct to Earth communication, there would be \textit{significant idle time on the surface of Europa}, which couldn't be utilized without autonomy \citep{europa_mission_paper_2022}. Finally, as with the Enceladus Orbilander concept, \acrshort{osia} would \textit{enable advanced instruments} (such microscopy videos for motility assessment) without significantly increasing downlink bandwidth or requiring new samples. For a mission as sample- and time-limited as Europa Lander, an additional biosignature capability could translate to a meaningful improvement in life detection sensitivity.}

\paragraph{Martian Rover}

\ed{The final reference mission we consider is a Martian rover (similar to Curiosity or Perseverance) focused on life detection. We include this reference to evaluate \acrshort{osia} benefits for a relatively close, high-bandwidth mission. Current rovers like Curiosity and Perseverance depend on orbiters (\acrshort{mro}, Odyssey, \acrshort{maven}, and \acrshort{tgo}), to support their data transmission, which ranges from 8 to \SI{2048}{\kilo\bit\per\second} \citep{gladden_2022}. For the sake of this estimation, we will consider an average data rate of \SI{1000}{\kilo\bit\per\second}.}

\ed{\acrshort{osia} inclusion on this future Martian rover concept would be driven by the need to support a daily operations planning cycle. Here, the mission cadence is likely to be similar to past rovers involving a sequence of detailed site explorations with short drives between each. While rover planning teams will not regularly face planning cycles as consequential as on Europa Lander, \textit{reaction time} is still a concern. The ability to rapidly return a few summary data products could guide mission teams to better optimize where and how they spend resources when searching for biosignatures. This reduces the risk that a sample site is abandoned prematurely or that valuable mission time is squandered at a site of low scientific value. \acrshort{osia} also offers the chance to \textit{collect many observations} and leverage \textit{soft retargeting} to analyze collected data over a long time period. Longer drives, for example, may leave substantial \textit{inactive time} for iterative \acrshort{osia} analysis of previously collected observations. As seen in the third row of Table \ref{tab:science_return_speed}, \acrshort{osia} with an existing MHz-scale processor would provide an \acrshort{oom} speedup over downlinking raw compressed products. This would be advantageous if the mission chose to collect data at high rates while at a site of interest, then process and transmit data during the upcoming drive. \acrshort{osia} also may also provide \textit{communication robustness} if there are any malfunctions in the orbital relay network, which would otherwise unexpectedly limit the rate and/or cadence of data transmission. Given the relatively high data bandwidth, a Mars mission may also be the most likely to deploy \textit{advanced instruments} like Raman or imaging spectrometers. Overall, \acrshort{osia} may offer some enhancements and risk reduction for a future rover concept, but the advantages are not as enabling compared to the Enceladus Orbilander and Europa Lander concepts exploring the outer solar system.}

\vspace{\baselineskip}

\ed{Ultimately, we seek to inform mission concept teams how processor choice and the presence or absence of \acrshort{osia} will impact science return. By comparing the combined runtime and transmission time of different strategies, Table \ref{tab:science_return_speed} displays the estimated \acrshort{oom} change in an observation's time-to-ground given choices in available data processing techniques (simple ZIP compression or \acrshort{osia}) and available compute platforms (MHz- or GHz-scale; Table \ref{tab:flightcompute}). At a high level, this analysis indicates that \acrshort{osia} has the greatest potential benefits for missions with low downlink bandwidth and a next-generation, Gigahertz-scale processor. For missions with high bandwidth, \acrshort{osia}'s potential benefits are diminished as the onboard runtime becomes a significant component of the total processing and transmission time. Regardless of the processor choice, we estimate that \acrshort{osia} will provide at least one \acrshort{oom} speedup in data return, which translates to either reduced time for a set data volume to reach ground teams or the ability to transmit a larger number of summarized observations in a fixed time window. Note that while we isolated these two aspects of the trade space for clarity, a true mission architecture study must explore a multitude of interconnected design choices simultaneously.}

\subsection{Path to Flight: Concept of Operations}

Given the fundamentally unknown nature of extraterrestrial life, the pursuit of biosignatures will require a \ed{nimble} interaction between science \ed{teams} and the \acrshort{osia}. The autonomy must therefore be capable of reconfiguration mid-mission to emphasize certain signals of interest and de-emphasize environmental distractions. This implies a new, more nimble \acrshort{conops} strategy than is typically used today. For existing \acrshort{osia} demonstrations such as \acrshort{aegis}, reconfiguration is treated as a rare event that requires extensive manual effort by the original research team, usually in response to data quality degradation. Moving forward, mission-enabling \acrshort{osia} will necessitate a more principled reconfiguration process; one that transparently captures new science intent, can be completed in hours or days, and produces reproducible, statistically defensible \acrshort{osia} configurations.

The defining novelty of \acrshort{osia} \acrshort{conops} is the specification of its goal-based behavior rather than manually-defined, imperative commands. In this regime, future operators will focus on selecting \acrshort{osia} configuration parameters that produce desirable outcomes for a statistically majority of the expected, future instrument observations. They will rely on new ground software tools to search this space of possible \acrshort{osia} configurations, likely observations, and their evaluated \acrshort{osia} outcomes to inform the selection of a single configuration that best matches the current science intent. This challenge and its solution are similar to mission formulation trade studies, where ensembles of simulations are routinely used to evaluate and select potential designs vs. predicted science outcomes. We are currently developing an open prototype of such a ground tool called the Data-driven Efficient Configuration of Instruments by Scientific Intent for Operational Needs (DECISION; \cite{Lightholder_DECISION_2023}), scheduled for completion in 2024.

\subsection{Early Integration of Science, Instrument, and Autonomy Teams}

\acrshort{osia} development is inherently interdisciplinary and therefore hinges on a strong collaboration between multiple stakeholders: the \acrshort{osia} developers themselves, science domain experts, instrument developers, and flight hardware and software teams. The practice of spiral development was used to facilitate this early interaction, which we describe with greater detail in \cite{Slingerland2022}. Initially, the autonomy team focused on a minimal end-to-end solution containing all the critical \acrshort{osia} modules as described in Figure \ref{fig:system_diagram}. After finishing a simple, working version of \acrshort{helm} and \acrshort{fame}, the team held biweekly feedback meetings with scientists and instrument developers to review results, identify outstanding challenges, and prioritize improvements in the hardware, algorithms, or observational data record. This accelerated development for all parties involved. For example, early \acrshort{mhi} and frame-to-frame pixel difference plots helped to identify and characterize instability in sample fluid flow (due to occasional clogs). In another case, the dropped frame validation check identified that extraneous background processes were occasionally overloading the flight computer early in F-Prime integration. These examples illustrate the advantages of integrating \acrshort{osia} early in development to increase system-level awareness and inform problem-solving across the mission. This approach is novel in the current mission environment where software and hardware teams are typically siloed until late-mission integration and testing. 

To engender trust with the science, instrument, and flight hardware teams, we deliberately regulated the complexity of the \acrshort{osia}'s underlying algorithms. Through repeated interaction and negotiation, we observed that simpler algorithms helped us better engage with the science team during spiral development, provide interpretable explanations of the \acrshort{osia}'s decisions, and meet timeliness requirements using the limited \ed{onboard} compute resources available. Specifically, the choice of the \acrshort{lap} tracking algorithm was influenced by existing science domain expert familiarity. Similarly, classical \acrshort{ml} models (e.g., \acrshortpl{rf} and \acrshortpl{gbt}) were selected for their explainability and relatively low compute requirements. While larger, more sophisticated algorithms such as \acrlongpl{cnn} or auto-encoders may bring certain performance advantages, they are also ``black-box'' models that are harder to interpret, require much more data to train, and use significantly more compute resources. Therefore, we chose not to rely on large or complex algorithms (such as deep learning models) as they would make for a poorly behaved citizen within the \acrshort{owls} onboard ecosystem. 

\subsection{Additional HELM and FAME Applications}

Beyond the planetary life detection context, \acrshort{helm} and \acrshort{fame} hold promise for many other use cases driven by their ability to systematically and efficiently process large numbers of microscopy observations \citep{Sweeney2019, Reimer_1997}. Some of these applications could also serve as useful opportunities to mature our systems for mission application. For example, automatically identifying motility styles could directly inform large-scale oceanic bacterial catalogs \citep{Grossart2001, Dinasquet2022, Mullen2020}, both to categorize known species and identify novel organisms. Low-cost, submersible \acrshortpl{dhm} are actively being developed for this purpose \citep{kent_2022}. For more direct societal impact, applications such as food safety analysis as well as beach and flood water safety are currently performed by manual motility inspection, as are the quantification of antibiotic efficacy and sepsis diagnoses \citep{Martin2016, Valderrama2015, Shahraki2019, Park2018, Tomenchok2020, Lazcka2007, Cholewiska2022, Saxena2014, Leonard2003}. Like our field trial use-case, the driving need in these terrestrial applications is guiding human attention to the most relevant data through summarization and prioritization. But whether at home or at interplanetary distances, limited by communication bandwidth or human attention, the most impactful role for \acrshort{osia} remains as a tool to help diagnose, discover, and understand the wealth of complex data that now surrounds us.

%% file: sections/7_appendix.tex
\newpage

\appendix
\section{Field Test Requirement Tables} \label{apx:reqs}
\newpage

\restartappendixnumbering

\global\pdfpageattr\expandafter{\the\pdfpageattr/Rotate 90}

\begin{longrotatetable}
\begin{deluxetable*}{l Y{3cm} Y{8cm} Y{8cm}}
\tablecaption{\acrshort{owls} field test level 1 and 2 requirements relevant to \acrshort{helm}/\acrshort{fame} \label{tab:reqs12}}
\tablehead{
    \colhead{Level} & \colhead{Short Name} & \colhead{Requirement} & \colhead{Justification}
}
\startdata
L1 & Autonomous Life Detection & 
The \acrshort{owls} suite shall autonomously investigate samples for molecular and cellular evidence of life in naturally occurring ocean-world analog environments. & 
Top level driving science capability.\\ 
\hline
L2 & Science Autonomy & 
The \acrshort{owls} Project shall summarize and prioritize observations from multiple samples for downlink to maximize returned evidence for life. & 
Limited data bandwidth and long communication delays prevent the return of most raw data and rapid ground-in-the-loop commanding.\\
\enddata
\end{deluxetable*}
\end{longrotatetable}


\begin{longrotatetable}
\begin{deluxetable*}{l Y{3cm} Y{8cm} Y{8cm} Y{1cm}}
\tablecaption{\acrshort{owls} \ed{field test} level 3 requirements \ed{relevant to \acrshort{helm}/\acrshort{fame}. The confirmation method ``Demo'' refers to the \acrshort{iadt} equivalent.} \label{tab:reqs3}}
\tablehead{
    \colhead{Level} & \colhead{Short Name} & \colhead{Requirement} & \colhead{Justification} & \colhead{Conf.}
}
\startdata
\ed{L3-1} & Data Summarization & 
The \acrshort{osia} shall produce reduced \ed{data volume science products} to characterize biosignatures while meeting the total baseline mission data budget. \ed{The data reduction ratio against the raw data volume shall be at least $1000 \times$}. & 
\ed{While we prescribe this requirement for the field test, observations in deep space mission concepts may be up to} 10,000 times larger than available downlink bandwidth. &
\ed{Demo \S\ref{sec:results_summarization}} \\
\hline
\ed{L3-2} & Data Prioritization & 
The \acrshort{osia} shall rank order science data products by their anticipated scientific value. & 
Missions may collect more observations than can be returned even with summarization capabilities, and data downlink may also be interrupted or delayed. Therefore, transmissions should include the most compelling products first at each opportunity. &
\ed{Demo \S\ref{sec:results_prio_tuning}}\\ 
\hline
\ed{L3-3} & Computation & 
The \acrshort{osia} shall complete all necessary validation, summarization, and prioritization using available flight computing resources within the acceptable timeliness window. & 
Flight computer capabilities and power budgets place significant restrictions on \acrshort{osia} algorithms. &
\ed{Demo L4-7} \\
\hline
\ed{L3-4} & Operational Monitoring & 
The \acrshort{osia} shall generate engineering telemetry sufficient to substantiate nominal operation of the autonomy and monitor instrument data quality. & 
The autonomy must support scientific conclusions with transparent, traceable behavior and provide operational parameters that can be trend-analyzed and tracked in the established mission operational paradigm. &
\ed{Demo \S\ref{sec:results_field}} \\ 
\hline
\ed{L3-5} & Reconfigurability & 
The \acrshort{osia}'s behavior shall be controlled by configuration files that may be updated during the mission to optimize its behavior. The \acrshort{osia} shall generate data products sufficient to detect the need for and inform reconfiguration. & 
The science operations team must have some awareness of non-prioritized observational contents to preserve the discovery of the unexpected as well as enable reconfiguration to pursue an evolving science focus or adapt to changing instrument behavior. &
\ed{Demo \S\ref{sec:results_field}} \\
\enddata
\end{deluxetable*}
\end{longrotatetable}


\begin{longrotatetable}
\begin{deluxetable*}{Y{1.5cm} Y{2.5cm} Y{8cm} Y{8cm} Y{1cm}}
\tablecaption{\acrshort{owls} \ed{field test} level 4 requirements \ed{relevant to \acrshort{helm}/\acrshort{fame}. The confirmation method ``Demo'' refers to the \acrshort{iadt} equivalent.} The requirements for each are nearly identical, save that \acrshort{fame} also includes absolute fluorescence intensity as a quantity of interest. \label{tab:reqs4}}
\tablehead{
    \colhead{ID} & \colhead{Short Name} & \colhead{Requirement} & \colhead{Justification} & \colhead{Conf.}
}
\startdata
\ed{L4-1 \textit{(L3-1,2)}} & Particle Tracking True Positives & 
\acrshort{helm} and \acrshort{fame} shall produce science data products that detect and track Well Resolved Targets and their motion within an Uncrowded Observation (raw image sequence vs. time) with at least 50\% True Track Coverage. & 
Capable particle tracking is needed for biosignature summarization and prioritization. Increasing sensitivity improves this metric but incurs more false positives. & 
\ed{Demo \S\ref{sec:results_tracking}}\\ \hline
\ed{L4-2 \textit{(L3-1,2)}} & Particle Tracking False Positives & 
\acrshort{helm} and \acrshort{fame} shall produce science data products that contain less than 10 False Track Points per Uncrowded Observation frame. & 
Performant particle tracking generates fewer false positive detections per frame to prevent crowding out of true biosignatures during downlink prioritization. Reducing sensitivity improves this metric but incurs more false negatives. & 
\ed{Demo \S\ref{sec:results_tracking}}\\ \hline
\ed{L4-3 \textit{(L3-1,2)}} & Motility Identification & 
\acrshort{helm} and \acrshort{fame} shall produce science data products that estimate the Empirical Probability of Life-Like Motility for each identified Well Resolved Target to inform summarization and prioritization. & 
The threshold for life-like detection should be interpretable to the science team, not arbitrary in units or meaning. & 
\ed{Demo \S\ref{sec:results_class}}\\ \hline
\ed{L4-4 \textit{(L3-1,2)}} & Fluorescence Identification & 
\acrshort{fame} shall produce science data products that track and characterize each Well Resolved Target in \acrshort{flfm} observations to inform summarization and prioritization. & 
Regardless of motility, information about any fluorescent particles (either innate or dye-induced) is valuable to the science team. & 
\ed{Demo \S\ref{sec:results_field}}\\ \hline
\ed{L4-5 \textit{(L3-2)}} & \acrfull{sue} & 
\acrshort{helm} and \acrshort{fame} shall produce a quantitative estimate of scientific utility for each observation. & 
By providing a single scalar estimate \ed{of} scientific utility, \acrshort{helm} and \acrshort{fame} enable transmission prioritization based on how likely science products from each observation are to fulfill the mission's science goals. & 
\ed{Demo \S\ref{sec:results_sue}}\\ \hline
\ed{L4-6 \textit{(L3-2)}} & \acrfull{dd} & 
\acrshort{helm} and \acrshort{fame} shall produce quantitative science data products that efficiently characterize unique aspects of each observation. & 
By providing a vector that describes the content of each observation, the \acrshort{osia} enables science teams to prioritize data in a manner that improves diversity through the inclusion of unique, unusual, or representative observations. & 
\ed{Demo \S\ref{sec:results_dd}}\\ \hline
\ed{L4-7 \textit{(L3-3)}} & \ed{Computation} & 
\ed{\acrshort{helm} and \acrshort{fame} shall produce science data products within an allocated compute time of ten (10) times the observation time.} &
\ed{The onboard autonomy must support field operations through timely summarization and prioritization of data despite limited computational resources.} &
\ed{Demo \S\ref{sec:results_compute}}\\ \hline
\ed{L4-8 \textit{(L3-4,5)}} & Background Summary Context & 
\acrshort{helm} and \acrshort{fame} shall produce science data products that summarize an entire observation including background context and data quality estimation as a function of time. & 
Background context is crucial to defend claims of life detection as well as recognize unanticipated observation contents, ensure proper system functioning, monitor instrument health, and support \acrshort{osia} reconfiguration. & 
\ed{Demo \S\ref{sec:results_valid}}\\ \hline
\ed{L4-9 \textit{(L3-4,5)}} & \ed{Logging} &
\ed{\acrshort{helm} and \acrshort{fame} shall generate a verbose log ensuring nominal operation and insight into data quality.} &
\ed{Logs are necessary to support efficient field explanation of autonomous system behavior. Logs also provide detailed records and explanations for observations that generated findings of high importance.} &
\ed{Demo \S\ref{sec:results_valid}}\\ \hline
\enddata
\end{deluxetable*}
\end{longrotatetable}

\global\pdfpageattr\expandafter{\the\pdfpageattr/Rotate 0}

\section{Acronyms and Abbreviations}

\printglossaries